\documentclass[10pt,aps,prd,twocolumn,showpacs,amsmath,amssymb,nofootinbib,eqsecnum,preprintnumbers,superscriptaddress]{revtex4-2}
\usepackage{amsmath,amssymb}
\usepackage{booktabs}   
\usepackage{multirow}   
\usepackage{array}      

\usepackage{amsmath,amssymb}
\usepackage{enumitem}  
\usepackage[usenames, dvipsnames]{color} 
\usepackage{graphicx} 
\usepackage{comment}
\usepackage[normalem]{ulem}
\usepackage[utf8]{inputenc}
\usepackage{url}
 \usepackage{xcolor}
 \usepackage{physics}

\usepackage{hyperref}  

\usepackage{graphicx}
\usepackage{dcolumn}
\usepackage{bm}

\newcommand{\be}{\begin{equation}}
\newcommand{\ee}{\end{equation}}
\newcommand{\ba}{\begin{eqnarray}}
\newcommand{\ea}{\end{eqnarray}}

\newcommand{\beq}{\begin{equation}}
\newcommand{\eeq}{\end{equation}}
\newcommand{\beqa}{\begin{eqnarray}}
\newcommand{\eeqa}{\end{eqnarray}}

\newcommand{\eps}{\varepsilon}

\newcommand{\A}[1]{A^{\!(#1)}}                 

\newcommand{\dg}{{{n}}}                           

\begin{document}


\title{From (Hidden) Symmetries to Stealth Solutions}





\author{David Kubiz\v n\'ak}
\email{david.kubiznak@matfyz.cuni.cz}
\affiliation{Institute of Theoretical Physics, Faculty of Mathematics and Physics,
Charles University, Prague, V Hole{\v s}ovi{\v c}k{\' a}ch 2, 180 00 Prague 8, Czech Republic}

\author{Robert B. Mann}
\email{rbmann@uwaterloo.ca}
\affiliation{%
Department of Physics and Astronomy, University of Waterloo, 
Waterloo, ON, N2L 3G1, Canada\\
Perimeter Institute  for Theoretical Physics, 31 Caroline Street North, Waterloo, ON, N2L 2Y5, Canada
}%

\author{Marek Mili{\v c}ka}
\email{marek.milicka@gmail.com }
\affiliation{Institute of Theoretical Physics, Faculty of Mathematics and Physics,
Charles University, Prague, V Hole{\v s}ovi{\v c}k{\' a}ch 2, 180 00 Prague 8, Czech Republic}

\date{June 5, 2026}

\begin{abstract}
In a recent 
paper, we have demonstrated that (conformal) Killing vectors 
give rise to 
stealth vector solutions of a 
specific bumblebee-type Proca
theory supplemented by fine tuned curvature terms. Here we show that such a  construction readily generalizes to hidden symmetries encoded in (conformal) Killing--Yano tensors, giving rise to the corresponding $p$-form stealth solutions. Similar to what happens with Killing vectors, 
the construction works on any background, 
providing a  `physical visualization' of its symmetries.
Several examples of 
spacetimes with so constructed $p$-form stealth hair are presented. 
\end{abstract}

 \maketitle

\section{Introduction}

Symmetries play a very important role in all areas of physics. Via Noether's theorem, they give rise to conserved quantities, linking conservation laws with special geometric properties of the system \cite{Cariglia:2014ysa}. 
In general relativity, symmetries are described by {\em Killing vectors} and their higher-rank generalizations known as (symmetric) {\em Killing tensors} \cite{staeckel1895integration, Walker:1970un} and (anti-symmetric) {\em Killing--Yano tensors} \cite{yano1952some, Penrose:1973um}. The latter two 
find their applications 
in  rotating black hole spacetimes, where they `substitute' for the lack of Killing vectors and stand behind many of the remarkable integrability and separability properties of such spacetimes, e.g. \cite{Frolov:2017kze}.

In a recent paper \cite{Kubiznak:2026vlq}, we have shown that (conformal) Killing fields give rise to {\em stealth} solutions of the following {\em Proca-type} theory for the vector field $A^a$:
\ba\label{LA} 
S_A&=&\int d^D\!x \sqrt{-g}{\cal L}_A\,,\nonumber\\
{\cal L}_A&=&-\frac{1}{4}F_{ab}F^{ab}+R_{ab}A^a A^b-\frac{D-1}{D}(\nabla_a A^a)^2\,,\qquad
\ea 
 reminiscent of what happens in {\em bumblebee}  gravity  \cite{Colladay:1998fq, Heisenberg:2014rta, Fan:2017bka}, 
where $D$ stands for the number of spacetime dimensions, $R_{ab}$ is the Ricci tensor, and the field strength of $A^a$ is denoted by   $F=dA$. Indeed, 
let $\xi^a$ be a conformal Killing vector, obeying the conformal Killing vector equation:
\be 
\nabla_{(a}\xi_{b)}=2\psi g_{ab}\,,\quad \psi=\frac{1}{D}\nabla_a\xi^a\,;
\ee 
it reduces to the standard Killing equation provided $\psi\nobreak=\nobreak0$.
Utilizing the corresponding integrability conditions:
\be 
\nabla_a\nabla_b\xi_c=\xi_d R^{d}{}_{abc}+\nabla_a(\psi g_{bc})+\nabla_b(\psi g_{ac})-\nabla_c(\psi g_{ab})\,,
\ee 
it can be shown \cite{Kubiznak:2026vlq} that
\be\label{Axi} 
A^a\equiv \xi^a
\ee 
has the following two properties:
\begin{enumerate}
    \item 
 It obeys the equations of motion derived from ${\cal L}_A$, \eqref{LA}: 
\be\label{EOMA} 
\nabla_a F^{ab}+2R^{ab}A_a+\frac{2(D-1)}{D}\nabla_b \nabla_a A^a=0\,.
\ee 
\item Such a solution is {\em stealth}, that is, its corresponding 
energy-momentum tensor $T^{ab}_A$, defined by (see \eqref{TabVector} for an explicit expression) 
\be 
\delta_g S_A
\equiv \frac{1}{2}\int d^D\!x \sqrt{-g}T_A^{ab}\delta g_{ab}\,,
\ee 
vanishes. 
\end{enumerate}

These two properties are purely {\em kinematic}, in that they are  valid on any background admitting the conformal Killing $\xi^a$, irrespective of the gravity field equations and the other field content, as long as the minimal coupling to ${\cal L}_A$ is assumed. 
In other words, we consider the following total Lagrangian:
\be\label{Stot} 
S_{\mbox{\tiny TOT}}=S_A+S_g+S_m=S_A+\int d^D\!x \sqrt{-g}\bigl({\cal L}_g+{\cal L}_m\bigr)\,,
\ee 
where 
$S_A$ is given by \eqref{LA}.  ${\cal L}_g$ describes the gravity part of the action; in Einstein gravity, this would simply be ${\cal L}_g=R/(2\kappa)$, but one can also add a cosmological constant or consider various modified gravity theories, such as Lovelock \cite{Lovelock:1971yv}.  ${\cal L}_m$ is the `other matter sector'; we assume minimal coupling to gravity and no interaction with the ${\cal L}_A$ sector.

The above construction provides, in some sense,  a `{\em visualization}' of Killing symmetries -- the corresponding vector fields are now treated as `physical' solutions of the above Proca-type theory, characterized by a special property that their gravitational backreaction vanishes. 
In this sense, for example, the flat space $D(D+1)/2$ Killing vectors provide a physical `reference frame' for the Minkowski spacetime.

At the same time, this construction generalizes, and to a certain extent elucidates, the well-known procedure for constructing {\em test Maxwell} fields in {\em vacuum} spacetimes from Killing symmetries  \cite{ Papapetrou:1966zz, Wald:1974np}. The main differences are that i) the identification \eqref{Axi} now works in any (not necessarily Ricci flat) spacetime, ii) the {\em test} Maxwell field is promoted to an  (arbitrarily) `{\em strong}' stealth Proca-type solution (with broken gauge symmetry), and iii) the construction also works for the  weaker structure of {\em conformal} Killing symmetries.

In fact, it is the aim of the current paper to show that the above construction can be extended to higher-form hidden symmetries, encoded in Killing--Yano tensors \cite{yano1952some, Penrose:1973um} and their conformal generalizations known as {\em conformal Killing--Yano (CKY) tensors} \cite{kashiwada1968conformal, tachibana1969conformal}. Namely, we will show that upon replacing $S_A$ in \eqref{Stot} with a specific, higher-form action 
\be\label{Somega}
S_\omega=\int d^D\!x\sqrt{-g}{\cal L}_\omega\,, 
\ee 
for a $p$-form $\omega$, the (conformal) Killing--Yano tensors automatically yield stealth solutions on any background admitting such symmetries. The corresponding action \eqref{Somega} then generalizes various Proca-type theories considered in the literature, such as bumblebee gravity \cite{Colladay:1998fq, Heisenberg:2014rta, Fan:2017bka}, to higher-form fields  coupled to curvature terms with fine tuned coupling constants.

The rest of our paper is organized as follows. To warm up, in the next section, we consider the case of a specific 2-form Proca-type theory $S_\omega$, \eqref{Somega}, and show that the Killing--Yano 2-forms give rise to its stealth solutions, providing also  two examples of the Kerr and the Taub-NUT spacetimes where such a construction applies.  The case of a general Killing--Yano $p$-form is treated in Sec.~\ref{pform}, with applications in higher-dimensional Kerr-NUT-AdS spacetimes. A possible generalization to conformal Killing--Yano tensors is discussed in Sec.~\ref{Sec4}. We conclude in Sec.~\ref{Sec5} with a summary and future directions. Appendix~\ref{AppA} contains details about the derivation of the energy-momentum tensor in the $p$-form theory studied in Secs.~\ref{pform} and \ref{Sec4}. Appendix~\ref{AppB} reviews the Plebanski--Demianski spacetimes \cite{Plebanski:1976gy} (and  their conformal-to-Carter generalizations \cite{Podolsky:2025tle, Ovcharenko:2025cpm, Gray:2025lwy}), together with their hidden symmetries that give rise to 2-form stealth solutions in these backgrounds.

\section{Killing--Yano 2-forms}\label{2form}

\subsection{2-form Proca-type theory}

Let us consider the following specific Proca-type action \eqref{Somega}: 
\be 
{\cal L}_\omega=-\frac{1}{6}F_{abc}F^{abc}+\frac{3}{2}R_{ac}\omega^{a}{}_d\omega^{cd}-\frac{3}{4}R_{abcd}\omega^{ab}\omega^{cd}\label{2formaction}\,
\ee 
for a 2-form $\omega_{ab}$, generalizing the action \eqref{LA} for (divergence-less) vector fields.
Here, $R_{abcd}$ is the Riemann tensor, and $F=d\omega$ is the `field strength' of $\omega$. 
The corresponding equations of motion for $\omega$ are then 
\be\label{EOMomega2} 
\nabla_a F^{abc}+3R^{[b}{}_a \omega^{|a|c]}-\frac{3}{2}R^{bc}{}_{ef}\omega^{ef}=0\,,
\ee 
and the associated energy momentum tensor $T^{ab}_\omega$, defined by 
\ba\label{Tabdef} 
\delta_g S_\omega\equiv\frac{1}{2}\int d^D\!x \sqrt{-g}T^{ab}_\omega\delta g_{ab}\,,
\ea 
now reads 
\begin{align}
T^{ab}_\omega =& g^{ab} {\cal L}_{\omega}+ F^a{}_{cd} F^{bcd}-6 R_{c}{}^{(a}\omega^{b)}{}_d \omega^{cd}\nonumber\\
&-3 R_{cd}\omega^{ca}\omega^{db}
+\frac{9}{2}R^{(a}{}_{cde}\omega^{b)c}\omega^{de}-\frac{3}{2}\nabla^2\bigl(\omega^{ac}\omega^b{}_c\bigr)
\nonumber\\
&+{3}\nabla^c \nabla^{(a}\bigl( \omega_{cd}\omega^{b)d}\bigr)
 -\frac{3}{2} g^{ab}\nabla_c \nabla_d \bigl(\omega^{ce}\omega^{d}{}_e\bigr) \nonumber\\
&+3\nabla_d \nabla_c \bigl(\omega^{ac}\omega^{bd}\bigr)\,,\label{p-formTab}
\end{align}
where 
details of the derivation (considering divergence-less forms $\omega$ and setting $p=2$) are
in Appendix~\ref{AppA}.

\subsection{Killing--Yano stealth solutions}

Let us now consider a Killing--Yano 2-form $f_{ab}$. It is defined by the following equations \cite{yano1952some, Penrose:1973um, Frolov:2017kze}:
\be 
f_{ab}=f_{[ab]}\,,\quad \nabla_{a}f_{bc}=\nabla_{[a}f_{bc]}=\frac{1}{3}(df)_{abc} 
\ee 
and obeys the following (integrability) conditions \cite{Frolov:2017kze}:
\ba \label{Integrability2form}
\nabla_c f^{ca}&=&0\,,\quad \nabla_{a}\nabla_b f_{cd}=\frac{3}{2}R_{ea[bc}f^e{}_{d]}\,,\nonumber\\
\nabla_{c}\nabla^{c}f_{ab} &=&\frac{1}{3}\nabla_c (df)^{cab}\nonumber\\
&=&  \frac{1}{2} R_{a}{}^{c} f_{bc} - \frac{1}{2} R_{b}{}^{c}f_{ac} + \frac{1}{2} R_{abcd} f^{cd}\,,\quad\\
R_{c(a}f^c{}_{b)}&=&0\,.\nonumber
\ea 
Importantly, upon identifying 
\be\label{omegaf} 
\omega_{ab}=f_{ab} 
\ee  
the middle equation is exactly the equation of motion \eqref{EOMomega2} for $\omega$.  
Thus, any Killing--Yano 2-form gives rise to solutions of the equation of motion \eqref{EOMomega2}.

To show that such a solution is also stealth takes some work. We have used  symbolic manipulation \cite{Garcia:2011} (see also the next section for a general analytic proof) to show that upon employing 
the equations of motion 
\eqref{EOMomega2} 
to eliminate the Riemann tensor and using the symmetries of the Killing--Yano tensor yields 
\be
T_f^{ab} =
\tfrac{3}{2} R^{cd} f^{a}{}_{c} f^{b}{}_{d} -  \tfrac{3}{2} R^{c(a} f^{b)d} f_{cd}
\ee
which is significantly simplified compared to
\eqref{p-formTab}.

Finally, using the last equation in \eqref{Integrability2form}, it is easy to see that the right hand side of the previous equation vanishes, yielding the stealth property
\be 
T^{ab}_f=0\,.
\ee 
So, every Killing--Yano 2-form yields a stealth solution on every background where it exists; note that this limits spacetimes to algebraic type D, or more special \cite{collinson1974existence, Mason:2010zzc}.

Note also that when evaluated on  solutions, the action \eqref{2formaction} 
 is `{\em holographic}', that is, it 
vanishes up to boundary terms. Indeed, we have 
\ba
{\cal L}_{\omega}&=& +\tfrac{1}{2}  \omega_{a b}\nabla_c (d\omega)^{cab}\nonumber\\
&& + \tfrac{3}{2} R_{ac} \omega^a{}_{b} \omega^{cb}- \tfrac{3}{4} R_{abcd} \omega^{ab} \omega^{cd}\nonumber \\
&=&\tfrac{1}{2} \omega_{ab}\left(\nabla_c (d\omega)^{cab}
 +3R^{a}{}_{c}\omega^{cb} - \tfrac{3}{2} R^{ab}{}_{cd}  \omega^{cd}\right)\nonumber\\
 &=&0\,,
\ea
upon integrating by parts and using  the antisymmetry of $\omega$ and \eqref{EOMomega2}. 
This is true in particular when restricting the action to Killing--Yano 2-forms, which are automatically solutions.
 It is also the same property that was observed in \cite{Kubiznak:2026vlq} for the action \eqref{LA} and its generic (conformal) Killing vector solutions.

\subsection{Canonical example: Kerr spacetime}

To provide an example of the above stealth fields, let us consider the Kerr solution written in the standard Boyer--Lindquist coordinates. The metric reads \cite{Kerr:1963ud} 
\ba\label{Kerr} 
ds^2&=&-\frac{\Delta}{\Sigma}(dt-a\sin^2\!\theta d\phi)^2+\frac{\Sigma}{\Delta}dr^2+\Sigma d\theta^2\nonumber\\
&&+\frac{\sin^2\!\theta}{\Sigma}(adt-[r^2+a^2]d\phi)^2\nonumber\\
&=&-e_0^2+e_1^2+e_2^2+e_3^2\,,
\ea 
where we have introduced the following tetrad:
\ba 
e_0&=&\sqrt{\frac{\Delta}{\Sigma}}(dt-a\sin^2\!\theta d\phi)\,,\quad e_2=\sqrt{\frac{\Sigma}{\Delta}}dr\,,\nonumber\\
e_1&=&\frac{\sin\theta}{\sqrt{\Sigma}}\Bigl(adt-(r^2+a^2)d\phi\Bigr)\,,
\quad e_3=\sqrt{\Sigma}d\theta\,,\qquad
\ea 
and the metric functions are given by 
\be 
\Sigma=r^2+a^2\cos^2\!\theta\,,\quad \Delta=r^2-2mr+a^2\,.
\ee

It is a vacuum solution of Einstein equations, $R_{ab}=0$, that apart from 
the explicit Killing fields:
\be 
k=\partial_t\,,\quad \eta=\partial_\phi\,,
\ee 
also admits a hidden symmetry encoded in the following Killing--Yano 2-form \cite{Penrose:1973um}: 
\ba\label{f4d} 
f&=&a \cos\theta dr\wedge (dt-a\sin^2\!\theta d\phi)\nonumber\\
&&-r\sin\theta d\theta\wedge \Bigl(a dt-(r^2+a^2)d\phi\Bigr)\nonumber\\
&=&a \cos \theta e_2\wedge e_0-r e_3\wedge e_1\,.
\ea 
The horizon is located at the largest root of $\Delta(r_+)=0$, that is $r_+=m+\sqrt{m^2-a^2}$, and it is a Killing horizon generated by Killing field 
\be 
\xi=k+\Omega_+ \eta\,,\quad \Omega_+=\frac{a}{r_+^2+a^2}\,,
\ee 
which becomes null there.

Focusing first on the Killing vectors, the following stealth solution, describing a  `charged' and `uniformly magnetized' Kerr black hole, was considered in \cite{Kubiznak:2026vlq}
\be 
A=-\frac{Q}{2m}k+\frac{1}{2}B(\eta+ak)\,, 
\ee 
where $Q$ and $B$ are 
the `stealth charges'.
It generalizes the test Maxwell field obtained in \cite{Wald:1974np}, and upon setting $B=0$ coincides with the Proca-type stealth solution recently obtained in \cite{Xu:2026zgd} by the Newman--Janis-type trick applied to a solution in bumblebee gravity. 
Focusing on this case, we have 
\ba\label{Aframe} 
A&=&\frac{-Q}{2m}\Bigl(-\sqrt{\frac{\Delta}{\Sigma}} e_0+\frac{a\sin\theta}{\sqrt{\Sigma}}e_1\Bigr)=-\frac{rQ}{\sqrt{\Sigma \Delta}}e_0+\frac{Q}{2m}dt\nonumber\\
&=&\Bigl(\frac{Q}{2m}-\frac{Qr}{\Sigma}\Bigr)dt-a\sin^2\!\theta\frac{rQ}{\Sigma} d\phi\,,
\ea 
and the stealth field can be assigned the following fluxes at infinity:
\be\label{fluxes} 
{\cal F}_A\equiv \frac{1}{4\pi}\int_{S^2_\infty} *F=Q\,,\quad \tilde {\cal F}_A\equiv \frac{1}{4\pi}\int_{S^2_\infty} F=0\,.
\ee 
As is obvious from the middle equality in \eqref{Aframe}, the stealth field $A$ is just a standard (singular on the horizon) Kerr--Newman Maxwell potential \cite{Newman:1965tw}, provided  the  $Q/(2m)dt$ term is dropped; in Maxwell theory this term is pure gauge.  Note, however, that it is precisely this term that  regularizes the solution on the  horizon, as seen from the first expression in \eqref{Aframe}, as the tetrad components remain finite there.

Turning next to the Kiling--Yano tensor \eqref{f4d}, we get the corresponding 2-form solution \eqref{omegaf}. In this case, however, the corresponding fluxes are over co-dimension 1 and co-dimension 3 hypersurfaces. It is not quite clear what such integrals would mean. For example, by analogy with the first integral in \eqref{fluxes}, we may define
\ba \label{Ff}
{\cal F}_f\equiv\frac{1}{2\pi} \int_{S^1}*F&=& -\frac{6amr\sin^2\!\theta}{\Sigma}\nonumber\\
&=&-\frac{6am\sin^2\!\theta}{r}+O(1/r^3)\,.\quad 
\ea 
where we have integrated over a circle at constant $t,r$, and  $\theta$. Although discoverable locally, such a `flux' vanishes at infinity.

\subsection{Another example: Taub-NUT spacetime}

As a second example, let us consider the Taub-NUT spacetime, given by \cite{taub1951empty, newman1963empty, Hawking:1998ct,  Clement:2015cxa}
\ba 
ds^2&=&-f(dt+2n \cos\theta d\phi)^2+\frac{dr^2}{f}\nonumber\\
&&\qquad +(r^2+n^2)(d\theta^2+\sin^2\!\theta d\phi)^2\,,
\ea 
where
\be 
f=\frac{r^2-2mr-n^2}{r^2+n^2}-\frac{3n^4-6n^2r^2-r^4}{\ell^2(r^2+n^2)}\,.
\ee 
Here,  $n$ is the NUT charge, $m$ is the mass parameter, and $\ell$ stands for the AdS radius.

The metric admits 4 Killing vectors corresponding to $SU(2) \times R$ isometry, see e.g. \cite{Gibbons:1987sp, Jezierski:2006fw} for a complete discussion of its  symmetries. Here, let us again focus on the following  two Killing fields
\be 
k=\partial_t\,,\quad \eta=\partial_\phi\,,
\ee 
together with a covariantly non-constant 
Killing--Yano 2-form $f$, which can be obtained as follows. Consider the following 1-form potential:
\be
b=-\frac{1}{2}r^2 dt - n(r^2+n^2)\cos\theta d\phi\,, 
\ee 
and its exterior derivative $h=db$. Then
\ba\label{KYTN} 
f&=&*db\nonumber\\
&=&ndr\wedge (dt+2n\cos\theta d\phi)\nonumber\\
&&\qquad -r\sin\theta(r^2+n^2)d\theta\wedge d\phi\,.
\ea

We may now consider the following stealth field:
\be 
A=-\frac{Q}{2m}k+\frac{1}{2}B\eta\,.
\ee 
Similar to the Kerr case, the second term has no effect on the fluxes \eqref{fintegrab}. At the same time, the first term now leads to divergent fluxes (diverging like $r^3$ and $r^2$, respectively). This is related to the well known fact that Komar charges in AdS diverge and have to be `renormalized'. To avoid this issue, let us consider the case of the  vanishing cosmological constant, setting $\ell\to \infty$. In that case, we find (upon sending $r\to\infty$)  
\be
{\cal F}_A=Q\,,\quad \tilde {\cal F}_A=-\frac{nQ}{m}\,.
\ee

We also have the 2-form stealth field $\omega$ identified with the Killing--Yano 2-form $f$, \eqref{KYTN}. The corresponding flux \eqref{Ff} again diverges in the AdS case. However, upon setting $\ell\to \infty$, we recover a finite result, given by  
\ba 
{\cal F}_f&=&
6nf\cos\theta\nonumber\\
&=&6n \cos\theta+O(1/r)\,,
\ea 
c.f. the quasi-local expression for Kerr \eqref{Ff}.

\section{Killing--Yano $p$-forms}\label{pform}

\subsection{Generalized $p$-form theory}

To generalize the above 2-form results, let us now consider the following $p$-form theory:
\ba\label{Proca_omega}
\mathcal{L}_\omega &=& -\tfrac{1}{2(p+1)} F_{c_1 \dots c_{p+1}} F^{c_1 \dots c_{p+1}}\nonumber\\
&& + \tfrac{(p+1)}{2} R_{ab}\Phi^{ab}- \tfrac{p^2-1}{4} R_{abcd}\Psi^{abcd}\,,
\ea
where, as always, $F=d\omega$, and we have introduced  
\ba 
\Phi^{ab}&\equiv&  \omega^a{}_{c_2 \dots c_p} \omega^{b c_2 \dots c_p}\,,\nonumber\\
\Psi^{abcd} &\equiv&\omega^{ab}{}_{c_3 \dots c_p} \omega^{cd c_3 \dots c_p}\,.
\ea 
Whereas the first object is automatically symmetric in its two indices: $\Phi^{ab}=\Phi^{(ab)}$, the latter has some `Riemann-type' symmetries, namely
\be\label{PsiRiem} 
\Psi^{abcd}=\Psi^{[ab]cd}=\Psi^{ab[cd]}=\Psi^{cdab}\,.    
\ee 
Note also that 
\be\label{psiphi} 
\Psi^{abc}{}_b=\Phi^{ac}=\Psi^{ba}{}_b{}^c\,.
\ee

The corresponding equations of motion for $\omega$ then read 
\ba\label{EOMpform}
    \nabla_a F^{a c_1\dots c_p} +(p+1)\!\!\!&&\!\!\! R^{a[c_1} \omega_{a}{}^{c_2 c_3 \dots c_p]} \nonumber\\
    -&&\!\!\!\! \tfrac{p^2-1}{2} R^{ab[c_1 c_2} \omega_{ab}{}^{ c_3 \dots c_p]}=0\,,\quad 
\ea
and, as shown in appendix~\ref{AppA} (see \eqref{TabfAppA}), the corresponding energy-momentum tensor is
\begin{align}\label{Tabf}
T^{ab}_\omega =& g^{ab} {\cal L}_\omega +F^{a}{}_{c_1 \dots c_p} F^{bc_1 \dots c_p}\nonumber\\
&-2(p+1)R^{c(a}\Phi^{b)}{}_c -(p^2-1) R_{cd}\Psi^{c(a|d|b)}\nonumber\\
&+\tfrac{3}{2}(p^2-1)R^{(a|cde|}\Psi^{b)}{}_{cde}\nonumber\\
&+ \tfrac{1}{2}(p-2)(p^2-1)R_{cdef}\omega^{acdc_4\dots c_p}\omega^{bef}{}_{c_4\dots c_p}\nonumber\\
&-\tfrac{1}{2}(p+1)\bigl(-2\nabla_c \nabla^{(a} \Phi^{b)c} +\nabla^2 \Phi^{ab} + g^{ab}\nabla_c \nabla_d \Phi^{cd}\bigr)\nonumber\\
&+(p^2-1)\nabla_d \nabla_c \Psi^{(a|c|b)d}\,.
\end{align}

We now want to prove that 
Killing--Yano tensors yield stealth solutions for the above theory. A Killing--Yano $p$-form $f$ is defined via the following two properties \cite{yano1952some, Penrose:1973um, Frolov:2017kze}:
\ba 
f_{a_1\dots a_p}&=&f_{[a_1\dots a_p]}\,,\nonumber\\
\nabla_a f_{a_1\dots a_p}&=&\nabla_{[a} f_{a_1\dots a_p]}=\frac{1}{p+1} (df)_{aa_1\dots a_p}\,.
\ea 
It follows that such objects obey the following integrability conditions (see, e.g.,  Appendix C in \cite{Frolov:2017kze}):
\ba\label{fintegrab} 
\nabla_a f^{aa_2\dots a_p}\!&=&\!0\,,\nonumber\\
\nabla^2 f_{aklc_4\dots c_p} \!&=&\!\tfrac{1}{p+1}\nabla^b (df)_{baklc_4\dots c_p}\nonumber\\
\!&=&\!- R_{ca}f^{c}{}_{klc_4\dots c_p} + \tfrac{p-1}{2}R_{dea[k}f^{de}{}_{lc_4\dots c_p]}\,,\quad \nonumber\\
R_{cd[kl}f_{|a|}{}^{cd}{}_{c_4\dots c_p]} \!&=&\! \tfrac{2}{p-2}(R_{e[k}f_{|a|}{}^e{}_{lc_4\dots c_p]}-R_{ea}f^e{}_{klc_4\dots c_p})\nonumber\\
\!&&\!+ R_{dea[k}f^{de}{}_{lc_4\dots c_p]}\,.
\ea 
The middle relation is precisely the equation of motion \eqref{EOMpform}. Thus, any Killing--Yano $p$-form is a solution of the field equation \eqref{EOMpform}.  Next, we want to show that such a solution is stealth.

\subsection{Proving stealth}

To show that for any Killing--Yano $p$-form $f$, the energy--momentum tensor \eqref{Tabf} vanishes, \be 
T^{ab}_f=0\,,
\ee 
we proceed in several steps.

\subsubsection{Preliminary steps}

{Let us start by noting that upon identifying $\omega$ with $f$, the `auxiliary tensors' $\Phi$ and $\Psi$ become the rank-2 Killing tensor and the `Riemann-type' generalized Killing tensor recently introduced in \cite{Arias:2016}, respectively; they now obey:
\be\label{Phi1} 
\nabla^{(c}\Phi^{ab)}=0\,,\quad \nabla_a \Phi^{ab}=-\tfrac{1}{2}\nabla^b \Phi^a{}_a\,,
\ee 
and 
\be\label{Phi2} 
\nabla^{(a}\Psi^{b|c|d)e}=0\,,\quad \nabla_a \Psi^{a(b|c|d)}=-\tfrac{1}{2}\nabla^c \Phi^{bd}\,,
\ee
where the last relation follows 
upon using \eqref{psiphi}. Similarly, we also have
\be\label{Phi3} 
\nabla_a \Psi^{(b|a|d)e}=\nabla^{(b}\Phi^{d)e}=-\tfrac{1}{2}\nabla^e \Phi^{bd}\,,
\ee 
where the last equality follows from the first relation \eqref{Phi1}.
}

\subsubsection{Terms proportional to metric}

Consider first the terms of $T^{ab}_f$ that are proportional to $g^{ab}$. These are
\be 
-\tfrac{p+1}{2}{\cal T}_1
\equiv{\cal L}_f-\tfrac{p+1}{2}\nabla_c\nabla_d \Phi^{cd} \,.
\ee 
Writing these explicitly in terms of $f$ and using the fact that Killing--Yano forms are divergence-less (first property in \eqref{fintegrab}), we get
\ba
{\cal T}_1&=&\nabla_cf_{a_1\dots a_p}\nabla^{c}f^{a_1\dots a_p} - R_{cd} f^{ca_2\dots a_p}f^d{}_{a_2\dots a_p} \nonumber\\
&&+ \tfrac{p-1}{2}R_{abcd}f^{aba_3\dots a_p}f^{cd}{}_{a_3\dots a_p} \nonumber\\
&&+ (\nabla_c \nabla_d f^{ca_2\dots a_p})f^d{}_{a_2\dots a_p} +  \nabla_d f^{ca_2\dots a_p} \nabla_c f^{da_2\dots a_p}\nonumber\\
&=& - R_{cd} f^{ca_2\dots a_p}f^d{}_{a_2\dots a_p} + \tfrac{p-1}{2}R_{abcd}f^{aba_3\dots a_p}f^{cd}{}_{a_3\dots a_p}\nonumber\\
&&+ (\nabla_c \nabla_d f^{ca_2\dots a_p})f^d{}_{a_2\dots a_p}\nonumber\\
&=&\Bigl(-\nabla^2 f_{da_2\dots a_p} - R_{c[d}f^c{}_{a_2\dots a_p]} \nonumber\\
&&\quad + \tfrac{p-1}{2}R_{ab[da_2 }f^{ab}{}_{a_3\dots a_p]}\Bigr)f^{da_2\dots a_p}\,,
\ea
where we used the Killing--Yano  condition that the gradient of $f$ is totally antisymmetric.
The terms in the parentheses are exactly the integrability condition for the Killing--Yano tensor (second relation in \eqref{fintegrab}), and so ${\cal L}_1$ vanishes, as expected. 

\subsubsection{Sorting out derivatives}

Let us next consider the terms of $T^{ab}_f$ that feature derivatives and no explicit Riemann tensor dependence.
These read 
\ba 
{\cal T}_2^{ab}&\equiv& 
F^{a}{}_{c_1 \dots c_p} F^{bc_1 \dots c_p}\nonumber\\
&&-\tfrac{1}{2}(p+1)\Bigl(-2\nabla_c \nabla^{(a} \Phi^{b)c} +\nabla^2 \Phi^{ab} \Bigr)\nonumber\\
&&+(p^2-1)\nabla_d \nabla_c \Psi^{(a|c|b)d}\,\nonumber\\
&=&
F^{a}{}_{c_1 \dots c_p} F^{bc_1 \dots c_p}-(p+1)\nabla^2\Phi^{ab}\nonumber\\
&&-\tfrac{1}{2}(p^2-1)\nabla^2 \Phi^{ab}\,\nonumber\\
&=&F^{a}{}_{c_1 \dots c_p} F^{bc_1 \dots c_p}-\tfrac{1}{2}(p+1)^2\nabla^2 \Phi^{ab}\,,
\ea 
where in the second equality, we have used \eqref{Phi2} and \eqref{Phi3}.

Writing out the auxiliary tensor in terms of $f$ then gives
\ba 
\nabla^2\Phi^{ab}&=&\tfrac{2}{(p+1)^2}F^{ac_1\dots c_p}F^b{}_{c_1\dots c_p}\nonumber\\
&&+\tfrac{2}{p+1}\nabla_c F^{c(a}{}_{c_2\dots c_p}f^{b)c_2\dots c_p}\,.
\ea 
Thus we have 
\be 
{\cal L}_2^{ab}=-(p+1)\nabla_c F^{c(a}{}_{c_2\dots c_p}f^{b)c_2\dots c_p}\,.
\ee

\subsubsection{Remaining terms}

We now want to show that the derivative terms cancel out the curvature terms. The latter read 
\ba 
{\cal T}_3^{ab}&\equiv&
-2(p+1)R^{c(a}\Phi^{b)}{}_c -(p^2-1) R_{cd}\Psi^{c(a|d|b)}\nonumber\\
&&+\tfrac{3}{2}(p^2-1)R^{(a|cde|}\Psi^{b)}{}_{cde}\\
&&+ \tfrac{1}{2}(p-2)(p^2-1)R_{cdef}f^{acdc_4\dots c_p}f^{bef}{}_{c_4\dots c_p}\,.\nonumber
\ea 
Applying the `curvature formula' \eqref{fintegrab} on the last term, we find 
\ba 
\tfrac{p-2}{2}R_{cdef}f^{acdc_4\dots c_p}f^{bef}{}_{c_4\dots c_p}\qquad\qquad\qquad\qquad\qquad\nonumber\\
=-
R^{c(a}\Phi^{b)}{}_c+R_{cd}\Psi^{c(a|d|b)}
+\tfrac{p-2}{2}R^{(a|cde|}\Psi^{b)}{}_{cde}\,.\quad
\ea 
Thus we have 
\ba 
{\cal T}_3^{ab}&\equiv&
-(p+1)^2R^{c(a}\Phi^{b)}{}_c \nonumber\\
&&+\tfrac{1}{2}(p^2-1)(p+1)R^{(a|cde|}\Psi^{b)}{}_{cde}\,.
\ea

Putting everything together, and upon expanding in terms of $f$, we thus have 
\ba 
T^{ab}_f &=&- (p+1)\nabla_c F^{c(a}{}_{c_2\dots c_p}f^{b)c_2\dots c_p}\nonumber\\
&&-(p+1)^2R^{c(a}f^{b)c_2\dots c_p}f_{cc_2\dots c_p} \nonumber\\
&&+\tfrac{1}{2}(p^2-1)(p+1)R^{(a}{}_{c_2de}f^{b)c_2 c_3\dots c_p}f^{de}{}_{c_3\dots c_p}\nonumber
\\
&=&-(p+1)\Bigl[\nabla_c F^{c(a}{}_{c_2\dots c_p}+(p+1)R^{c(a}f_{cc_2\dots c_p} \nonumber\\
&&\quad 
-\tfrac{p^2-1}{2}R^{(a}{}_{c_2de}f^{|de|}{}_{c_3\dots c_p}
\Bigr]f^{b)c_2\dots c_p}\,.
\ea
Since the bracket is precisely the integrability condition \eqref{fintegrab}, it vanishes, proving that the field is stealth.

\subsection{Example: Kerr-NUT-AdS spacetime in higher dimensions}

It is well-known that the existence of Killing--Yano tensors severely restricts the admissible spacetimes. For example, a Killing--Yano 2-form can only exist in Weyl type D (or algebraically more special) spacetimes, see \cite{collinson1974existence} and \cite{Mason:2010zzc} for a proof in four and higher dimensions.

In four dimensions, a general spacetime admitting a rank-3 Killing--Yano tensor has been constructed \cite{dietz1981space}. Such spacetimes admit a (Hodge dual)  closed conformal Killing vector $k$, $f=*k$, and in particular, include metrics with covariantly constant Killing vector $k=\partial_v$:
\be 
ds^2=2dudv+g_{ij}(u,x,y)dx^i dx^j\,.
\ee

Here, we turn to a (perhaps more interesting) example of spacetimes with higher-rank Killing--Yano tensors -- the (off-shell)  {\em Kerr-NUT-AdS spacetime}. Parameterizing the spacetime dimension ${D=2\dg+\eps}$ (with $\eps=0$ in even and $\eps=1$ in odd dimensions), the metric reads \cite{Frolov:2017kze}:
\ba\label{KerrNUTAdSmetric}
ds^2&=&\sum_{\mu=1}^\dg\;\biggl[\; \frac{U_\mu}{X_\mu}\,{dx_{\mu}^{2}}
  +\, \frac{X_\mu}{U_\mu}\,\Bigl(\,\sum_{j=0}^{\dg-1} \A{j}_{\mu}d\psi_j \Bigr)^{\!2}
  \;\biggr]\nonumber\\
  &&  +\eps\frac{c}{\A{\dg}}\Bigl(\sum_{k=0}^\dg \A{k}d\psi_k\!\Bigr)^{\!2}\;,
\ea
where 
the functions ${\A{k}}$, ${\A{j}_\mu}$, and ${U_\mu}$ are `symmetric polynomials' of coordinates ${x_\mu}$:
\ba\label{AUdefs} \A{k}&=&\!\!\!\!\!\sum_{\substack{\nu_1,\dots,\nu_k=1\\\nu_1<\dots<\nu_k}}^\dg\!\!\!\!\!x^2_{\nu_1}\dots x^2_{\nu_k}\;,
\qquad
  \A{j}_{\mu}=\!\!\!\!\!\sum_{\substack{\nu_1,\dots,\nu_j=1\\\nu_1<\dots<\nu_j\\\nu_i\ne\mu}}^\dg\!\!\!\!\!x^2_{\nu_1}\dots x^2_{\nu_j}\;,\nonumber\\
U_{\mu}&=&\prod_{\substack{\nu=1\\\nu\ne\mu}}^\dg(x_{\nu}^2-x_{\mu}^2)\;,
\ea
$c$ is a constant that appears in odd dimensions, 
and each metric function ${X_\mu}$ is a function of a single coordinate ${x_\mu}$:
\begin{equation}\label{Xfcdependence}
    X_\mu=X_\mu(x_\mu)\;.
\end{equation}
When the vacuum Einstein equations with a cosmological constant are imposed, 
$X_\mu$ takes the following explicit form:
\be 
X_\mu=
\begin{cases}
-2b_\mu x_\mu+\sum_{k=0}^n c_k (x_\mu)^{2k}\quad D\ \mbox{is even}\\
-\frac{c}{x_\mu^2}-2b_\mu+\sum_{k=1}^n c_k (x_\mu)^{2k}\quad D\ \mbox{is odd}\,.
\end{cases}
\ee 
Interestingly, the  described (hidden) symmetries (see below) are present irrespective of the field equations, and exist for a general class of the off shell spacetimes characterized by \eqref{Xfcdependence}.

The metric admits a \textit{closed conformal Killing--Yano 2-form} $h$,  expressed as the exterior derivative of a one-form potential $b$:
\be 
h=db\,,\quad 
b=\frac{1}{2}\sum_{k=0}^{n-1} A^{(k+1)} d\psi_k\,.\label{PKYb}
\ee 
This object generates a tower of closed conformal Killing--Yano $(2p)$-forms, given by 
\be 
h_1\equiv h\,,\quad h_p\equiv \underbrace{h\wedge \dots\wedge h}_{\mbox{\small $p$-times}}\,, 
\ee 
and their Hodge duals are the rank-$(D-2p)$ Killing--Yano tensors
\be 
f_{p}=*h_p\,.
\ee 
It is these objects that give raise to stealth $(D-2p)$-form solutions $\omega_p=f_p$ of the above generalized Proca theory \eqref{Proca_omega}.

Of course, the above example includes, as a special case (see \cite{Frolov:2017kze} for details), the higher-dimensional multiply-spinning black holes of Myers and Perry \cite{Myers:1986un}, penetrated by the corresponding stealth solutions. We leave the calculation of the corresponding stealth field fluxes and their potential thermodynamic imprints to future studies.

\section{Conformal Killing--Yano conjecture}\label{Sec4}


It also seems possible to generalize the above considerations to a weaker structure of CKY tensors. 
Namely, let us consider the following generalization of the action \eqref{Proca_omega}:
\ba\label{ProcaCKY}
\mathcal{L}_\omega &\equiv& -\tfrac{1}{2(p+1)} F_{c_1 \dots c_{p+1}}F^{c_1 \dots c_{p+1}}\nonumber\\
&&- \tfrac{(D-p+1)(D-p)(p+1)}{2p^2} \xi_{c_2\dots c_p}\xi^{c_2\dots c_p}\nonumber\\
&& + \tfrac{p+1}{2} R_{ab}\Phi^{ab} - \tfrac{p^2-1}{4} R_{abcd} \Psi^{abcd}\,,\label{CKYaction}
\ea
where, as before $F=d\omega$,
\be \label{xidef}
\xi_{c_2\dots c_p}\equiv \frac{p}{D-p+1}\nabla^a \omega_{ac_2\dots c_p}\,,
\ee 
and $\Phi$ and $\Psi$ are defined via \eqref{psiphi}.

The corresponding equations of motion read:
\begin{align}\label{CKYEOM}
    \nabla^a F_{a c_1\dots c_p} =& -(p+1) R_{e[c_1} \omega^{e}{}_{c_2 c_3 \dots c_p]} \nonumber\\
    &+ \tfrac{p^2-1}{2} R_{ed[c_1 c_2} \omega^{ed}{}_{ c_3 \dots c_p]} \nonumber\\
    &- \tfrac{(p+1)(D-p)}{p} \nabla_{[c_1} \xi_{c_2 \dots c_p]}\,,
\end{align}
and the energy-momentum tensor is given by
\begin{align}\label{TabofCKY}
T^{ab}_\omega =& g^{ab} {\cal L}_\omega +F^{a}{}_{c_1 \dots c_p} F^{bc_1 \dots c_p}\nonumber\\
&-2(p+1)R^{c(a}\Phi^{b)}{}_c -(p^2-1) R_{cd}\Psi^{c(a|d|b)}\nonumber\\
&+\tfrac{3}{2}(p^2-1)R^{(a|cde|}\Psi^{b)}{}_{cde}\nonumber\\
&+ \tfrac{1}{2}(p-2)(p^2-1)R_{cdef}\omega^{acdc_4\dots c_p}\omega^{bef}{}_{c_4\dots c_p}\nonumber\\
&-\tfrac{1}{2}(p+1)\bigl(-2\nabla_c \nabla^{(a} \Phi^{b)c} +\nabla^2 \Phi^{ab} + g^{ab}\nabla_c \nabla_d \Phi^{cd}\bigr)\nonumber\\
&+(p^2-1)\nabla_d \nabla_c \Psi^{(a|c|b)d}\,\nonumber\\
&+\tfrac{(D-p)(p+1)}{p}\Bigl[-2\omega_{(a|c_2\dots c_p}\nabla_{b)}\xi^{c_2\dots c_p} \nonumber\\
    &\qquad -\tfrac{(p-1)(D-p+1)}{p}\xi_{(a|c_3\dots c_p|}\xi_{b)}{}^{c_3\dots c_p}\nonumber\\
    &\qquad -2(p-1)\omega_{d(a|c_3\dots c_p|} \nabla^d\xi_{b)}{}^{c_3\dots c_p}\nonumber\\    &\qquad +g_{ab}\nabla^e(\omega_{ec_2\dots c_p}\xi^{c_2\dots c_p})\Bigr]\,,
\end{align}
where the last four $\xi$ terms (and the $\xi$ term implicitly hidden in extended ${\cal L}_\omega$) give the extension of \eqref{Tabf}, see derivation in Appendix \ref{AppA}.

Consider now a generalization of Killing--Yano tensors to the CKY case \cite{kashiwada1968conformal, tachibana1969conformal}. 
These are $p$-forms $k$, obeying 
\be
\nabla_a k_{b_1\dots b_p} = \nabla_{[a}k_{b_1\dots b_p]} + \frac{p}{D-p+1} g_{a[b_1}\nabla^c k_{|c|b_2\dots b_p]}\,.
\ee
Taking a divergence of this equation and using the identity  `$d = \nabla \wedge$' for the exterior derivative yields
\be
\nabla^2 k_{b_1\dots b_p} = \tfrac{1}{p+1}\nabla^{a} (dk)_{a b_1\dots b_p} + \tfrac{p}{D-p+1}\nabla_{[b_1}\nabla^c k_{|c|b_2\dots b_p]}\,.
\ee
We now use the Weitzenböck identity 
\be
\nabla^2 k_{b_1\dots b_p} = \nabla^a (dk)_{a b_1\dots b_p} + p \nabla_{[b_1}\nabla^c k_{|c|b_2\dots b_p]} + Wk_{b_1\dots b_p}\,,
\ee
where  $W$ is the Weitzenböck operator acting on a $p$-form as
\be\label{Ws}
W \sigma_{c_1 c_2 c_3 \dots c_p} =p R_{e[c_1} \sigma^{e}{}_{c_2 c_3 \dots c_p]} - \tfrac{p(p-1)}{2} R_{ed[c_1 c_2} \sigma^{ed}{}_{ c_3 \dots c_p]}\,. 
\ee
Re-arranging the terms, we get the equation
\begin{align}\label{CKYIntCond}
     \nabla^a (dk)_{a b_1\dots b_p} =& - \tfrac{p+1}{p} Wk_{b_1\dots b_p}\nonumber\\
     &- \tfrac{(p+1)(D-p)}{D-p+1} \nabla_{[b_1}\nabla^c k_{|c|b_2\dots b_p]}\,,
\end{align}
which, upon using \eqref{Ws} yields precisely the equation of motion \eqref{CKYEOM}.
Thus, CKY forms yield solutions of the above theory \eqref{ProcaCKY}. 

While we have not explicitly checked that such solutions are also stealth, we believe that this is the case, leaving the corresponding  calculation for a future study.  We have, however, explicitly checked that this is the case for the four-dimensional Plebanski--Demianski spacetimes  \cite{Plebanski:1976gy} (and their conformal-to-Carter generalizations \cite{Podolsky:2025tle, Ovcharenko:2025cpm, Gray:2025lwy}) and their two CKY 2-forms;  see Appendix~\ref{AppB}.

Let us also note that the action \eqref{CKYaction} is again holographic on solutions, similar to what happens with the action \eqref{Somega}.
Indeed, up to boundary terms we have 
\ba
\mathcal{L}_\omega &=& \tfrac{1}{2}\omega_{abc_3\dotsc_p}\nabla_e(d\omega)^{eabc_3\dotsc_p}\nonumber\\
    &&+ \tfrac{(D-p)(p+1)}{2p^2}\omega_{abc_3\dots c_p}\nabla^a\xi^{bc_3\dots c_p}\nonumber\\
    &&+\tfrac{p+1}{2}R_{ab}\Phi^{ab}-\tfrac{p^2-1}{4}R_{abcd}\Psi^{abcd}\nonumber\\
    &=&\tfrac{1}{2}\omega^{abc_3\dots c_p}\Bigl(\nabla^e(d\omega)_{eabc_3\dots c_p}\nonumber\\
    &&\quad +\tfrac{(D-p)(p+1)}{D-p+1}\nabla_{[a}\nabla^e\omega_{|e|bc_3\dots c_p]}\nonumber\\
    &&\quad +(p+1)R_{e[a}\omega^e{}_{bc_3\dots c_p]}\nonumber\\
    &&\quad - \tfrac{p^2-1}{2}R_{cd[ab}\omega^{cd}{}_{c_3\dots c_p]}\Bigr)\,,
\ea
which vanishes due to the equation of motion \eqref{CKYEOM}\,.

\section{Conclusions}\label{Sec5}

In this paper, we have demonstrated that hidden symmetries encoded in Killing--Yano tensors give rise to stealth $p$-form solutions of the properly generalized Proca-type actions with curvature terms. This generalizes the recent results on (conformal) Killing vectors obtained in  \cite{Kubiznak:2026vlq}. We also expect it   to work for a weaker structure of  conformal Killing--Yano symmetries.  
To support this conjecture, we have explicitly shown that it applies to the Plebanski--Demianski spacetimes \cite{Plebanski:1976gy}  and their special cases, as well as to the recently obtained Ovcharenko--Podolsky class of metrics \cite{Podolsky:2025tle, Ovcharenko:2025cpm, Gray:2025lwy} (see Appendix~\ref{AppB}).
If true in general, the above Kerr-NUT-AdS spacetimes \eqref{KerrNUTAdSmetric}, as well as their conformal generalizations given by $\Omega^{-2}ds^2$,  provide examples of spacetimes with conformal Killing--Yano symmetries where such a  construction can  immediately be applied.

The generalized Proca-type actions constructed above, namely \eqref{Somega}, \eqref{Proca_omega}, and  \eqref{ProcaCKY}, are not gauge invariant. However, gauge invariance can easily be restored by applying the Stueckelberg trick. For example, instead of the action \eqref{Somega}, we may consider 
\ba\label{Stuck2form} 
{\cal L}_\omega&=&-\frac{1}{6}(d\omega)_{abc}(d\omega)^{abc}\nonumber\\
&&+\frac{3}{2}R^a{}_c(\omega_{ad}-\nabla_{[a}\phi_{d]})(\omega^{cd}-\nabla^{[c}\phi^{d]})\nonumber\\
&&-\frac{3}{4}R_{abcd}(\omega^{ab}-\nabla^{[a}\phi^{b]})(\omega^{cd}-\nabla^{[c}\phi^{d]})\,,
\ea
where we have introduced an `auxiliary' vector field $\phi^a$, and the action is now gauge invariant under 
\be
\omega_{ab} \to \omega_{ab} - \nabla_{[a}\alpha_{b]} 
\qquad \phi_a \to \phi_a + \alpha_a\,,
\ee 
with a gauge `potential' $\alpha_a$. A similar trick, of course, also applies to higher-form actions 
\eqref{Proca_omega} and  \eqref{ProcaCKY}.

Let us also stress that although in this paper we have concentrated on `standard' Killing--Yano forms, it seems likely that a similar construction also applies to their `torsion generalizations' \cite{Kubiznak:2009qi}, which naturally find applications in supergravity and string theory backgrounds and are no longer limited to type D spacetimes, e.g. \cite{Houri:2010qc, Houri:2010fr}. It also remains to be seen whether the above construction can be generalized to (totally symmetric) Killing tensors \cite{staeckel1895integration, Walker:1970un} and/or (mixed symmetry) generalized Killing tensors of \cite{collinson2000generalized, Arias:2016}. The 
integrability conditions for the former were recently studied in \cite{Houri:2017tlk}. 
 If so, this would provide a universal  `physical field realization' of symmetries in gravitational theories.

\subsection*{Acknowledgements}
D.K. acknowledges support from the Charles University Research Center Grant No. UNCE24/SCI/016. 
This work was supported in part by the Natural Sciences and Engineering Research Council of Canada.

\appendix

\section{Derivation of the p-form stealth energy-momentum tensor
}\label{AppA}

The energy-momentum tensor for the action \eqref{LA} was derived  in \cite{Kubiznak:2026vlq}. It reads 
\ba\label{TabVector} 
T^{ab}_A&=&g^{ab}{\cal L}_A+F^{ac}F^b{}_c-4A^{(a}A_c R^{b)c}-\nabla^2 (A^aA^b)\nonumber\\
&&-g^{ab} \nabla_c \nabla_d(A^c A^d)+2\nabla_c\nabla^{(a}(A^{|c|}A^{b)})\nonumber\\
&&-\tfrac{2(D-1)}{D} g^{ab} A^{c} \nabla_{c}\nabla_{d}A^{d} \,,
\ea 
where ${\cal L}_A$ is given by \eqref{LA}.
In this appendix, we 
provide a detailed derivation of a generalization of this energy-momentum tensor for a $p$-form action \eqref{ProcaCKY}. 
This is defined via \eqref{Tabdef}, that is
\be 
\delta_g S_\omega\equiv\frac{1}{2}\int d^D\!x \sqrt{-g}T^{ab}_\omega\delta g_{ab}\,.
\ee   
We do the calculation in several steps: {First, we employ the well-known identities 
\be\label{Varids} 
\delta \sqrt{-g}=-\tfrac{1}{2}\sqrt{-g}g_{ab}\delta g^{ab}\,,\quad \delta g_{ab}=-g_{ac}g_{bd}\delta g^{cd}\,,
\ee 
to vary the determinant in the action. This simply yields a term 
\be 
T^{(0)}_{ab}=g_{ab}{\cal L}_\omega\,.
\ee 
Second,} we will perform the variation of the  contractions in ${\cal L}_\omega$; this defines the algebraic part $T_{ab}^{\mbox{\tiny alg}}$. 
Third, we vary the curvature terms, employing the Palatini identity, obtaining $T_{ab}^{\mbox{\tiny Palatini}}$. Finally, we look at the divergence terms, obtaining $T_{ab}^{\mbox{\tiny div}}$. 
During the derivation we use the fact that the exterior derivative is independent of the connection, and hence its variation vanishes.

\subsubsection{Algebraic part}

The first term, $-\tfrac{1}{2(p+1)} (d\omega)_{a c_1 \dots c_p} (d\omega)^{a c_1 \dots c_p}$, contains $p+1$ equivalent contractions, the variation is therefore
\be
-\tfrac{1}{2} (d\omega)_{a c_1 \dots c_p} (d\omega)_b{}^{c_1 \dots c_p}\,.
\ee

The Ricci term $\tfrac{p+1}{2}R_{ab} \Phi^{ab}$ has two contractions between the tensors and $p-1$ contractions inside $\Phi$. Since both tensors are symmetric, the contractions between them yield the same contribution, so the variation is
\be
(p+1)R_{c(a}\Phi_{b)}{}^c + \tfrac{p^2-1}{2} R_{cd}\Psi^c{}_{(a}{}^d{}_{b)}\,.
\ee

In the last term, $-\tfrac{p^2-1}{4}R^a{}_{bcd}\Psi_a{}^{bcd}$, we use the fact that $\Psi$ has $(p-2)$ internal contractions. Moreover, $\Psi$ has symmetries of the Riemann tensor, namely \eqref{PsiRiem}, and so the contraction for the three raised indices yields the same contribution. In the end, the variation is
\ba
-\tfrac{3(p^2-1)}{4}R_{(a|cde|}\Psi_{b)}{}^{cde} \qquad \qquad \qquad\qquad \qquad \nonumber\\
- \tfrac{(p-2)(p^2-1)}{4}R_{cdef}\omega^{cd}{}_{ac_4\dots c_p}\omega^{ef}{}_b{}^{c_4\dots c_p}\,.
\ea
Putting everything together,  
$T_{ab}^{\mbox{\tiny alg}}$ is then minus twice the sum of these terms, that is, 
\begin{align}
T_{ab}^{\mbox{\tiny alg}}=& F_{a c_1 \dots c_p} F_b{}^{c_1 \dots c_p}-2 (p+1)R_{c(a}\Phi_{b)}^c\nonumber\\
& -(p^2-1) R_{cd}\Psi^c{}_{(a}{}^d{}_{b)}+\tfrac{3(p^2-1)}{2}R_{(a|cde|}\Psi_{b)}{}^{cde}\nonumber\\
&+ \tfrac{(p-2)(p^2-1)}{2}R_{cdef}\omega^{cd}{}_{ac_4\dots c_p}\omega^{ef}{}_b{}^{c_4\dots c_p}\,.
\end{align}

\subsubsection{Palatini term}
Let us next vary the curvature terms, returning first to the term $\tfrac{p+1}{2}R_{ab}\Phi^{ab}$, and taking the variation of the Ricci tensor, for which we use the {\em Palatini identity:}
\be
\delta(R_{ab}) =   \nabla_e \delta\Gamma_{a}{}^e{}_{b} -\nabla_b \delta \Gamma_{e}{}^e{}_{a}\,.
\ee
Integrating by parts, and throwing away the boundary (total divergence) terms, we then have 
\ba
\tfrac{p+1}{2}\delta R_{ab}\Phi^{ab}&=& -\tfrac{p+1}{2}(\nabla_e\Phi^{ab} - \nabla_c \Phi^{ac}\delta^b_e) \delta \Gamma_a {}^e{}_b\nonumber\\
&=&-\tfrac{p+1}{4}\nabla^e\Phi^{ab}(\nabla_a \delta g_{eb} + \nabla_b \delta g_{ae}\nonumber\\
&&-\nabla_e\delta g_{ab})
+ \tfrac{p+1}{4}\nabla_c\Phi^{ac} g^{be} \nabla_a \delta g_{eb},\qquad\ 
\ea
where we used the fact that 
\be
\delta\Gamma_a{}^b{}_c = \tfrac{1}{2}g^{bd}(\nabla_a \delta g_{dc} + \nabla_c \delta g_{ad} - \nabla_d \delta g_{ca})\,,
\ee
together with the symmetry of $g^{be}$.
Exchanging the variation of the metric for the variation of its inverse gives a minus sign; see \eqref{Varids}. Thus, we have 
\begin{align}
\tfrac{p+1}{2}\delta R_{ab}\Phi^{ab}=&
\tfrac{p+1}{4}(\nabla_a \nabla^e\Phi^{ab}\delta g_{eb} + \nabla_b \nabla^e\Phi^{ab}\delta g_{ae} \nonumber\\
&- \nabla_e \nabla^e\Phi^{ab}\delta g_{ab}) - \tfrac{p+1}{4} \nabla_a \nabla_c\Phi^{ac} g^{be}\delta g_{eb}\nonumber\\
=& -\tfrac{p+1}{4}(\nabla^c \nabla_a \Phi_{cb} +\nabla^c \nabla_b \Phi_{ac}\nonumber\\
&-\nabla^2 \Phi_{ab} - \nabla_c \nabla_d \Phi^{cd} g_{ab})\delta g^{ab}\,.\label{RicciVariation}
\end{align}

For the second term, $-\tfrac{p^2-1}{4} R^a{}_{bcd}\Psi_a{}^{bcd}$, we will use the following Palatini identity:
\be
\delta R^a{}_{bcd} = \nabla_c \delta\Gamma_d{}^a{}_b - \nabla_d \delta\Gamma_c{}^a{}_b\,.
\ee
Using the symmetry of $\Psi$, we then have
\begin{align}
-\tfrac{p^2-1}{4}\delta R^a{}_{bcd}\Psi_a{}^{bcd}=& -\tfrac{p^2-1}{2}(\nabla_c\delta\Gamma_d{}^a{}_b)\Psi_a{}^{bcd} \nonumber\\
=& \tfrac{p^2-1}{2}\delta\Gamma_d{}^a{}_b\nabla_c\Psi_a{}^{bcd}\nonumber\\
=&\tfrac{p^2-1}{4}(\nabla_b \delta g_{ed} + \nabla_d \delta g_{be}\nonumber\\
& \quad- \nabla_e \delta g_{db}) \nabla_c\Psi^{ebcd}\,.
\end{align}
Since $\Psi$ is antisymmetric in the first two indices, the second term does not contribute. 
By integrating by parts, we then get
\begin{align}
\tfrac{1-p^2}{4}\delta R^a{}_{bcd}\Psi_a{}^{bcd}&=
\tfrac{1-p^2}{4}(\nabla_b \nabla_c \Psi^{ebcd} \delta g_{de} \nonumber\\
&\quad - \nabla_e \nabla_c \Psi^{ebcd}\delta g_{db})\nonumber\\
&=\tfrac{p^2-1}{4}(\nabla_d \nabla_c \Psi_b{}^{dc}{}_a - \nabla_d \nabla_c \Psi^d{}_b{}^c{}_a)\delta g^{ab}\nonumber\\
&=\tfrac{1-p^2}{2}(\nabla_d \nabla_c \Psi_{(a}{}^c{}_{b)}{}^d)\delta g^{ab}\,,
\end{align}
where we used that $\Psi$ has the same symmetries as the Riemann tensor in the last step.
The contribution to the energy--momentum tensor is minus twice the sum of the contributions of the curvature terms, that is
\begin{align}
T_{ab}^{\mbox{\tiny Palatini}}=& -\tfrac{p+1}{2}(-\nabla^c \nabla_a \Phi_{cb} -\nabla^c \nabla_b \Phi_{ac}\nonumber\\
&+\nabla^2 \Phi_{ab} + \nabla_c \nabla_d \Phi^{cd} g_{ab})\nonumber\\
&+ (p^2-1)\nabla_d \nabla_c \Psi_{(a}{}^c{}_{b)}{}^d\,.
\end{align}
Thus, the total energy-momentum tensor without the divergence parts, that is for the action \eqref{Proca_omega}, reads: 
\begin{align}\label{TabfAppA}
T^{ab}_\omega =& g^{ab} {\cal L}_\omega +F^{a}{}_{c_1 \dots c_p} F^{bc_1 \dots c_p}\nonumber\\
&-2(p+1)R^{c(a}\Phi^{b)}{}_c -(p^2-1) R_{cd}\Psi^{c(a|d|b)}\nonumber\\
&+\tfrac{3}{2}(p^2-1)R^{(a|cde|}\Psi^{b)}{}_{cde}\nonumber\\
&+ \tfrac{1}{2}(p-2)(p^2-1)R_{cdef}\omega^{acdc_4\dots c_p}\omega^{bef}{}_{c_4\dots c_p}\nonumber\\
&-\tfrac{1}{2}(p+1)\bigl(-2\nabla_c \nabla^{(a} \Phi^{b)c} +\nabla^2 \Phi^{ab} + g^{ab}\nabla_c \nabla_d \Phi^{cd}\bigr)\nonumber\\
&+(p^2-1)\nabla_d \nabla_c \Psi^{(a|c|b)d}\,.
\end{align}
This is the energy-momentum tensor used in Sec.~\ref{pform} of the main text.

\subsubsection{Divergence additions}
The more general action \eqref{CKYaction}, studied in Sec.~\ref{Sec4}, has the following additional divergence term:  
\begin{align}
    - \tfrac{(D-p+1)(D-p)(p+1)}{2p^2} \xi_{c_2\dots c_p}\xi^{c_2\dots c_p} \qquad\qquad  \nonumber\\
    =-\tfrac{(D-p)(p+1)}{2p}\nabla^c\omega_{cc_2\dots c_p}\xi^{c_2\dots c_p}\,.
\end{align}
It provides extra terms to the energy momentum tensor through modified $\mathcal{L}_\omega$ and through an explicit contribution $T^{\mbox{\tiny div}}_{ab}$.
To shorten notation, let us look at the variation of the term $\nabla^c\omega_{cc_2\dots c_p}\xi^{c_2\dots c_p}$ without the prefactor.

The algebraic variation, in the same sense as above, gives
\begin{equation}
 2\nabla_{(a}\omega_{b)c_2\dots c_p}\xi^{c_2\dots c_p} +(p-1)\nabla^e\omega_{e(a|c_2\dots c_p|}\xi_{b)}{}^{c_2\dots c_p}\,,   
\end{equation}
and only the variation of the derivatives remains.
That is 
\begin{align}
    2(\delta\nabla^c)\omega_{cc_2\dots c_p}\xi^{c_2\dots c_p} =& - 2g^{cd}(\delta\Gamma_c{}^e{}_d)\omega_{ec_2\dots c_p}\xi^{c_2\dots c_p}\nonumber\\
    &-2g^{cd}\sum^p_{i=2}\delta\Gamma_c{}^e{}_{c_i} \omega_{dc_2\dots e \dots c_p}\xi^{c_2\dots c_p}\,.
\end{align}
Since
\begin{align}
g^{cd}\delta\Gamma_c{}^e{}_d=&\tfrac{1}{2}g^{cd}g^{ef}(\nabla_c\delta g_{df}+\nabla_d\delta g_{cf}-\nabla_f\delta g_{dc})\nonumber\\
    =&g^{ef}\nabla^d\delta g_{df} - \tfrac{1}{2}g^{cd}\nabla^e\delta g_{cd}
\end{align}
the first term, after throwing away the boundary terms and raising the variation indices, contributes as
\begin{equation}
    t_1 \equiv \delta g^{ab} [-2\nabla_{(a}(\omega_{b)c_2\dots c_p}\xi^{c_2\dots c_p})+g_{ab}\nabla^e(\omega_{ec_3\dots c_p}\xi^{c_3\dots c_p})]\,.
\end{equation}

One summand of the second term is
\begin{align}
    -g^{cd}g^{ef}(\nabla_c\delta g_{c_if}+\nabla_{c_i}\delta g_{cf}-\nabla_f\delta g_{c_ic})\omega_{d c_2\dots e\dots c_p} \xi^{c_2\dots c_p}\,.
\end{align}
We notice that due to the antisymmetry of $\omega$ in indices $[de]$, the terms in the parentheses are antisymmetrized in indices $[cf]$. The middle term vanishes due to the symmetry of the metric variation, and the remaining two terms contribute equally. 
Due to the antisymmetry of $\omega$ and $\xi$, the contributions of all terms in the sum are identical; integrating by parts, dropping boundary terms, and raising variation indices, we obtain the final contribution
\begin{equation}
    t_2=\delta g^{ab}[-2(p-1) \nabla^d(\omega_{d(a|c_3\dots c_p|}\xi_{b)}{}^{c_3\dots c_p})]\,.
\end{equation}

The extra terms are minus twice the variation, that is,
\begin{align}\label{Textra}
    T_{ab}^{\mbox{\tiny div}} =& \tfrac{(D-p)(p+1)}{p}\left[ 2\nabla_{(a}\omega_{b)c_2\dots c_p}\xi^{c_2\dots c_p}\right.\nonumber\\
    &+(p-1)\nabla^e\omega_{e(a|c_3\dots c_p|}\xi_{b)}{}^{c_3\dots c_p}\nonumber\\
    &-2\nabla_{(a}(\omega_{b)c_2\dots c_p}\xi^{c_2\dots c_p})+g_{ab}\nabla^e(\omega_{ec_2\dots c_p}\xi^{c_2\dots c_p})\nonumber\\
    &\left.-2(p-1) \nabla^d(\omega_{d(a|c_3\dots c_p|}\xi_{b)}{}^{c_3\dots c_p})\right]\nonumber\\
    =&\tfrac{(D-p)(p+1)}{p}\left[-2\omega_{(a|c_2\dots c_p}\nabla_{b)}\xi^{c_2\dots c_p} \right.\nonumber\\
    &-(p-1)\nabla^e\omega_{e(a|c_3\dots c_p|}\xi_{b)}{}^{c_3\dots c_p}\nonumber\\
    &-2(p-1)\omega_{d(a|c_3\dots c_p|} \nabla^d\xi_{b)}{}^{c_3\dots c_p}\nonumber\\
    &+g_{ab}\nabla^e(\omega_{ec_2\dots c_p}\xi^{c_2\dots c_p})\left.\right]\,.
\end{align} 

Putting everything together, we arrive at the total energy-momentum tensor:
\begin{align}\label{TabofCKYApp}
T^{ab}_\omega =& g^{ab} {\cal L}_\omega +F^{a}{}_{c_1 \dots c_p} F^{bc_1 \dots c_p}\nonumber\\
&-2(p+1)R^{c(a}\Phi^{b)}{}_c -(p^2-1) R_{cd}\Psi^{c(a|d|b)}\nonumber\\
&+\tfrac{3}{2}(p^2-1)R^{(a|cde|}\Psi^{b)}{}_{cde}\nonumber\\
&+ \tfrac{1}{2}(p-2)(p^2-1)R_{cdef}\omega^{acdc_4\dots c_p}\omega^{bef}{}_{c_4\dots c_p}\nonumber\\
&-\tfrac{1}{2}(p+1)\bigl(-2\nabla_c \nabla^{(a} \Phi^{b)c} +\nabla^2 \Phi^{ab} + g^{ab}\nabla_c \nabla_d \Phi^{cd}\bigr)\nonumber\\
&+(p^2-1)\nabla_d \nabla_c \Psi^{(a|c|b)d}\,\nonumber\\
&+\tfrac{(D-p)(p+1)}{p}\Bigl[-2\omega_{(a|c_2\dots c_p}\nabla_{b)}\xi^{c_2\dots c_p} \nonumber\\
    &\qquad -\tfrac{(p-1)(D-p+1)}{p}\xi_{(a|c_3\dots c_p|}\xi_{b)}{}^{c_3\dots c_p}\nonumber\\
    &\qquad -2(p-1)\omega_{d(a|c_3\dots c_p|} \nabla^d\xi_{b)}{}^{c_3\dots c_p}\nonumber\\    &\qquad +g_{ab}\nabla^e(\omega_{ec_2\dots c_p}\xi^{c_2\dots c_p})\Bigr]\,,
\end{align}
used in Sec.~\ref{Sec4}.

\section{Plebanski--Demianski spacetimes}\label{AppB}

The Plebanski--Demianski spacetime \cite{Plebanski:1976gy} is the most general type D spacetime with aligned electromagnetic field. The corresponding metric reads 
\be 
g=\frac{1}{\Omega^2}g^{\mbox{\tiny (C)}}\,,
\ee 
where $g^{\mbox{\tiny (C)}}$ is the (off-shell) Carter's metric element
\cite{carter1968new, carter1968hamilton}
 \ba 
g^{\mbox{\tiny (C)}}&=&-\frac{\Delta_r}{\Sigma}(dt+y^2d\phi)^2+\frac{\Sigma}{\Delta_r} dr^2\nonumber\\
&&+\frac{\Delta_y}{\Sigma}(dt-r^2d\phi)^2+\frac{\Sigma}{\Delta_y}dy^2\,,
\ea 
where $\Sigma=r^2+y^2$, and the conformal factor is given by 
\be 
\Omega=1-yr\,.
\ee 
To provide a solution of the Einstein--Maxwell-$\Lambda$ equations of motion, the two metric functions $\Delta_r$ and $\Delta_y$ take the following specific form: 
\ba
\Delta_r&=&k+{\rm e}^2+{\rm g}^2-2mr+\epsilon r^2-2nr^3-(k+\Lambda/3)r^4\;,\nonumber\\
\Delta_y&=&k+2ny-\epsilon y^2+2my^3-(k+{\rm e}^2+{\rm g}^2+\Lambda/3)y^4\,.\nonumber\\
\ea
Constants $k, m, \epsilon, n$ are free parameters that are related to mass, rotation, NUT parameter, and acceleration, and ${\rm e}$ and ${\rm g}$ are the electromagnetic charges of the aligned Maxwell field given by 
\be 
A=-\frac{{\rm e} r}{\Sigma}(dt+y^2d\phi)-\frac{{\rm g}y}{\Sigma}(dt-r^2d\phi)\,.
\ee 
We refer to \cite{Griffiths:2005qp} for a discussion and the interpretation of special cases of the Pleba\'nski--Demia\'nski metric; see also \cite{Ovcharenko:2024yyu, Ovcharenko:2025fxg} for more recent developments.

The above metric admits two (Hodge dual) CKY 2-forms. These are given by, e.g. \cite{Kubiznak:2007kh}: 
\be 
h=\frac{1}{\Omega^3}db\,,\quad b=\frac{1}{2}\bigl([y^2-r^2)dt-r^2y^2 d\phi\bigr)\,,
\ee 
or more explicitly 
\be 
h=\frac{1}{\Omega^3}\Bigl(ydy\wedge (dt-r^2d\phi)-rdr\wedge (dt+y^2d\phi)\Bigr)\,,
\ee 
and its Hodge dual 
\ba 
k&=&*h\nonumber\\
&=&\frac{1}{\Omega^3}\bigl(ydr\wedge (dt+y^2d\phi)+rdy\wedge (dt-r^2d\phi)\bigr)\,.\quad 
\ea 
It can be shown that both of these tensors remain CKY tensors also for the off-shell spacetimes, that is, for arbitrary 
\be 
\Delta_r=\Delta_r(r)\,,\quad \Delta_y=\Delta_y(y)\,,
\ee 
and any 
\be 
\Omega=\Omega(r,y,t,\phi)\,,
\ee 
that is for general off-shell conformal-to-Carter spacetimes.
We have explicitly checked that even in this general off-shell case, they give rise to the stealth 2-form solutions of the theory \eqref{ProcaCKY}. 
In particular, this is true for the recently discovered  Ovcharenko--Podolsky type D spacetimes with non-aligned electromagnetic fields \cite{Podolsky:2025tle, Ovcharenko:2025cpm, Gray:2025lwy}.


\begin{thebibliography}{45}%
\makeatletter
\providecommand \@ifxundefined [1]{%
 \@ifx{#1\undefined}
}%
\providecommand \@ifnum [1]{%
 \ifnum #1\expandafter \@firstoftwo
 \else \expandafter \@secondoftwo
 \fi
}%
\providecommand \@ifx [1]{%
 \ifx #1\expandafter \@firstoftwo
 \else \expandafter \@secondoftwo
 \fi
}%
\providecommand \natexlab [1]{#1}%
\providecommand \enquote  [1]{``#1''}%
\providecommand \bibnamefont  [1]{#1}%
\providecommand \bibfnamefont [1]{#1}%
\providecommand \citenamefont [1]{#1}%
\providecommand \href@noop [0]{\@secondoftwo}%
\providecommand \href [0]{\begingroup \@sanitize@url \@href}%
\providecommand \@href[1]{\@@startlink{#1}\@@href}%
\providecommand \@@href[1]{\endgroup#1\@@endlink}%
\providecommand \@sanitize@url [0]{\catcode `\\12\catcode `\$12\catcode `\&12\catcode `\#12\catcode `\^12\catcode `\_12\catcode `\%12\relax}%
\providecommand \@@startlink[1]{}%
\providecommand \@@endlink[0]{}%
\providecommand \url  [0]{\begingroup\@sanitize@url \@url }%
\providecommand \@url [1]{\endgroup\@href {#1}{\urlprefix }}%
\providecommand \urlprefix  [0]{URL }%
\providecommand \Eprint [0]{\href }%
\providecommand \doibase [0]{https://doi.org/}%
\providecommand \selectlanguage [0]{\@gobble}%
\providecommand \bibinfo  [0]{\@secondoftwo}%
\providecommand \bibfield  [0]{\@secondoftwo}%
\providecommand \translation [1]{[#1]}%
\providecommand \BibitemOpen [0]{}%
\providecommand \bibitemStop [0]{}%
\providecommand \bibitemNoStop [0]{.\EOS\space}%
\providecommand \EOS [0]{\spacefactor3000\relax}%
\providecommand \BibitemShut  [1]{\csname bibitem#1\endcsname}%
\let\auto@bib@innerbib\@empty
\bibitem [{\citenamefont {Cariglia}(2014)}]{Cariglia:2014ysa}%
  \BibitemOpen
  \bibfield  {author} {\bibinfo {author} {\bibfnamefont {M.}~\bibnamefont {Cariglia}},\ }\bibfield  {title} {\bibinfo {title} {{Hidden Symmetries of Dynamics in Classical and Quantum Physics}},\ }\href {https://doi.org/10.1103/RevModPhys.86.1283} {\bibfield  {journal} {\bibinfo  {journal} {Rev. Mod. Phys.}\ }\textbf {\bibinfo {volume} {86}},\ \bibinfo {pages} {1283} (\bibinfo {year} {2014})},\ \Eprint {https://arxiv.org/abs/1411.1262} {arXiv:1411.1262 [math-ph]} \BibitemShut {NoStop}%
\bibitem [{\citenamefont {Staeckel}(1895)}]{staeckel1895integration}%
  \BibitemOpen
  \bibfield  {author} {\bibinfo {author} {\bibfnamefont {P.}~\bibnamefont {Staeckel}},\ }\href@noop {} {\emph {\bibinfo {title} {Sur l'integration de l'{\'e}quation diff{\`e}rentielle de Hamilton}}}\ (\bibinfo  {publisher} {Gauthier-Villars},\ \bibinfo {year} {1895})\BibitemShut {NoStop}%
\bibitem [{\citenamefont {Walker}\ and\ \citenamefont {Penrose}(1970)}]{Walker:1970un}%
  \BibitemOpen
  \bibfield  {author} {\bibinfo {author} {\bibfnamefont {M.}~\bibnamefont {Walker}}\ and\ \bibinfo {author} {\bibfnamefont {R.}~\bibnamefont {Penrose}},\ }\bibfield  {title} {\bibinfo {title} {{On quadratic first integrals of the geodesic equations for type [22] spacetimes}},\ }\href {https://doi.org/10.1007/BF01649445} {\bibfield  {journal} {\bibinfo  {journal} {Commun. Math. Phys.}\ }\textbf {\bibinfo {volume} {18}},\ \bibinfo {pages} {265} (\bibinfo {year} {1970})}\BibitemShut {NoStop}%
\bibitem [{\citenamefont {Yano}(1952)}]{yano1952some}%
  \BibitemOpen
  \bibfield  {author} {\bibinfo {author} {\bibfnamefont {K.}~\bibnamefont {Yano}},\ }\bibfield  {title} {\bibinfo {title} {Some remarks on tensor fields and curvature},\ }\href@noop {} {\bibfield  {journal} {\bibinfo  {journal} {Annals of Mathematics}\ }\textbf {\bibinfo {volume} {55}},\ \bibinfo {pages} {328} (\bibinfo {year} {1952})}\BibitemShut {NoStop}%
\bibitem [{\citenamefont {Penrose}(1973)}]{Penrose:1973um}%
  \BibitemOpen
  \bibfield  {author} {\bibinfo {author} {\bibfnamefont {R.}~\bibnamefont {Penrose}},\ }\bibfield  {title} {\bibinfo {title} {{Naked singularities}},\ }\href {https://doi.org/10.1111/j.1749-6632.1973.tb41447.x} {\bibfield  {journal} {\bibinfo  {journal} {Annals N. Y. Acad. Sci.}\ }\textbf {\bibinfo {volume} {224}},\ \bibinfo {pages} {125} (\bibinfo {year} {1973})}\BibitemShut {NoStop}%
\bibitem [{\citenamefont {Frolov}\ \emph {et~al.}(2017)\citenamefont {Frolov}, \citenamefont {Krtous},\ and\ \citenamefont {Kubiznak}}]{Frolov:2017kze}%
  \BibitemOpen
  \bibfield  {author} {\bibinfo {author} {\bibfnamefont {V.~P.}\ \bibnamefont {Frolov}}, \bibinfo {author} {\bibfnamefont {P.}~\bibnamefont {Krtous}},\ and\ \bibinfo {author} {\bibfnamefont {D.}~\bibnamefont {Kubiznak}},\ }\bibfield  {title} {\bibinfo {title} {{Black holes, hidden symmetries, and complete integrability}},\ }\href {https://doi.org/10.1007/s41114-017-0009-9} {\bibfield  {journal} {\bibinfo  {journal} {Living Rev. Rel.}\ }\textbf {\bibinfo {volume} {20}},\ \bibinfo {pages} {6} (\bibinfo {year} {2017})},\ \Eprint {https://arxiv.org/abs/1705.05482} {arXiv:1705.05482 [gr-qc]} \BibitemShut {NoStop}%
\bibitem [{\citenamefont {Kubiznak}\ \emph {et~al.}(2026)\citenamefont {Kubiznak}, \citenamefont {Mann},\ and\ \citenamefont {Mili{\v{c}}ka}}]{Kubiznak:2026vlq}%
  \BibitemOpen
  \bibfield  {author} {\bibinfo {author} {\bibfnamefont {D.}~\bibnamefont {Kubiznak}}, \bibinfo {author} {\bibfnamefont {R.~B.}\ \bibnamefont {Mann}},\ and\ \bibinfo {author} {\bibfnamefont {M.}~\bibnamefont {Mili{\v{c}}ka}},\ }\bibfield  {title} {\bibinfo {title} {{Proca-type Hair of Rotating Black Holes in Higher Dimensions}},\ }\href@noop {} {\  (\bibinfo {year} {2026})},\ \Eprint {https://arxiv.org/abs/2605.23077} {arXiv:2605.23077 [hep-th]} \BibitemShut {NoStop}%
\bibitem [{\citenamefont {Colladay}\ and\ \citenamefont {Kostelecky}(1998)}]{Colladay:1998fq}%
  \BibitemOpen
  \bibfield  {author} {\bibinfo {author} {\bibfnamefont {D.}~\bibnamefont {Colladay}}\ and\ \bibinfo {author} {\bibfnamefont {V.~A.}\ \bibnamefont {Kostelecky}},\ }\bibfield  {title} {\bibinfo {title} {{Lorentz violating extension of the standard model}},\ }\href {https://doi.org/10.1103/PhysRevD.58.116002} {\bibfield  {journal} {\bibinfo  {journal} {Phys. Rev. D}\ }\textbf {\bibinfo {volume} {58}},\ \bibinfo {pages} {116002} (\bibinfo {year} {1998})},\ \Eprint {https://arxiv.org/abs/hep-ph/9809521} {arXiv:hep-ph/9809521} \BibitemShut {NoStop}%
\bibitem [{\citenamefont {Heisenberg}(2014)}]{Heisenberg:2014rta}%
  \BibitemOpen
  \bibfield  {author} {\bibinfo {author} {\bibfnamefont {L.}~\bibnamefont {Heisenberg}},\ }\bibfield  {title} {\bibinfo {title} {{Generalization of the Proca Action}},\ }\href {https://doi.org/10.1088/1475-7516/2014/05/015} {\bibfield  {journal} {\bibinfo  {journal} {JCAP}\ }\textbf {\bibinfo {volume} {05}},\ \bibinfo {pages} {015}},\ \Eprint {https://arxiv.org/abs/1402.7026} {arXiv:1402.7026 [hep-th]} \BibitemShut {NoStop}%
\bibitem [{\citenamefont {Fan}(2018)}]{Fan:2017bka}%
  \BibitemOpen
  \bibfield  {author} {\bibinfo {author} {\bibfnamefont {Z.-Y.}\ \bibnamefont {Fan}},\ }\bibfield  {title} {\bibinfo {title} {{Black holes in vector-tensor theories and their thermodynamics}},\ }\href {https://doi.org/10.1140/epjc/s10052-018-5540-7} {\bibfield  {journal} {\bibinfo  {journal} {Eur. Phys. J. C}\ }\textbf {\bibinfo {volume} {78}},\ \bibinfo {pages} {65} (\bibinfo {year} {2018})},\ \Eprint {https://arxiv.org/abs/1709.04392} {arXiv:1709.04392 [hep-th]} \BibitemShut {NoStop}%
\bibitem [{\citenamefont {Lovelock}(1971)}]{Lovelock:1971yv}%
  \BibitemOpen
  \bibfield  {author} {\bibinfo {author} {\bibfnamefont {D.}~\bibnamefont {Lovelock}},\ }\bibfield  {title} {\bibinfo {title} {{The Einstein tensor and its generalizations}},\ }\href {https://doi.org/10.1063/1.1665613} {\bibfield  {journal} {\bibinfo  {journal} {J. Math. Phys.}\ }\textbf {\bibinfo {volume} {12}},\ \bibinfo {pages} {498} (\bibinfo {year} {1971})}\BibitemShut {NoStop}%
\bibitem [{\citenamefont {Papapetrou}(1966)}]{Papapetrou:1966zz}%
  \BibitemOpen
  \bibfield  {author} {\bibinfo {author} {\bibfnamefont {A.}~\bibnamefont {Papapetrou}},\ }\bibfield  {title} {\bibinfo {title} {{Champs gravitationnels stationnaires {\`a} sym{\'e}trie axiale}},\ }\href@noop {} {\bibfield  {journal} {\bibinfo  {journal} {Ann. Inst. H. Poincare Phys. Theor. A}\ }\textbf {\bibinfo {volume} {4}},\ \bibinfo {pages} {83} (\bibinfo {year} {1966})}\BibitemShut {NoStop}%
\bibitem [{\citenamefont {Wald}(1974)}]{Wald:1974np}%
  \BibitemOpen
  \bibfield  {author} {\bibinfo {author} {\bibfnamefont {R.~M.}\ \bibnamefont {Wald}},\ }\bibfield  {title} {\bibinfo {title} {{Black hole in a uniform magnetic field}},\ }\href {https://doi.org/10.1103/PhysRevD.10.1680} {\bibfield  {journal} {\bibinfo  {journal} {Phys. Rev. D}\ }\textbf {\bibinfo {volume} {10}},\ \bibinfo {pages} {1680} (\bibinfo {year} {1974})}\BibitemShut {NoStop}%
\bibitem [{\citenamefont {Kashiwada}(1968)}]{kashiwada1968conformal}%
  \BibitemOpen
  \bibfield  {author} {\bibinfo {author} {\bibfnamefont {T.}~\bibnamefont {Kashiwada}},\ }\bibfield  {title} {\bibinfo {title} {On conformal killing tensor},\ }\href@noop {} {\bibfield  {journal} {\bibinfo  {journal} {Natural Science Report}\ }\textbf {\bibinfo {volume} {19}},\ \bibinfo {pages} {68} (\bibinfo {year} {1968})}\BibitemShut {NoStop}%
\bibitem [{\citenamefont {Tachibana}(1969)}]{tachibana1969conformal}%
  \BibitemOpen
  \bibfield  {author} {\bibinfo {author} {\bibfnamefont {S.-i.}\ \bibnamefont {Tachibana}},\ }\bibfield  {title} {\bibinfo {title} {On conformal killing tensor in a riemannian space},\ }\href@noop {} {\bibfield  {journal} {\bibinfo  {journal} {Tohoku Mathematical Journal, Second Series}\ }\textbf {\bibinfo {volume} {21}},\ \bibinfo {pages} {56} (\bibinfo {year} {1969})}\BibitemShut {NoStop}%
\bibitem [{\citenamefont {Plebanski}\ and\ \citenamefont {Demianski}(1976)}]{Plebanski:1976gy}%
  \BibitemOpen
  \bibfield  {author} {\bibinfo {author} {\bibfnamefont {J.~F.}\ \bibnamefont {Plebanski}}\ and\ \bibinfo {author} {\bibfnamefont {M.}~\bibnamefont {Demianski}},\ }\bibfield  {title} {\bibinfo {title} {{Rotating, charged, and uniformly accelerating mass in general relativity}},\ }\href {https://doi.org/10.1016/0003-4916(76)90240-2} {\bibfield  {journal} {\bibinfo  {journal} {Annals Phys.}\ }\textbf {\bibinfo {volume} {98}},\ \bibinfo {pages} {98} (\bibinfo {year} {1976})}\BibitemShut {NoStop}%
\bibitem [{\citenamefont {Podolsky}\ and\ \citenamefont {Ovcharenko}(2025)}]{Podolsky:2025tle}%
  \BibitemOpen
  \bibfield  {author} {\bibinfo {author} {\bibfnamefont {J.}~\bibnamefont {Podolsky}}\ and\ \bibinfo {author} {\bibfnamefont {H.}~\bibnamefont {Ovcharenko}},\ }\bibfield  {title} {\bibinfo {title} {{Kerr Black Hole in a Uniform Bertotti-Robinson Magnetic Field: An Exact Solution}},\ }\href {https://doi.org/10.1103/rfgv-ybz5} {\bibfield  {journal} {\bibinfo  {journal} {Phys. Rev. Lett.}\ }\textbf {\bibinfo {volume} {135}},\ \bibinfo {pages} {181401} (\bibinfo {year} {2025})},\ \Eprint {https://arxiv.org/abs/2507.05199} {arXiv:2507.05199 [gr-qc]} \BibitemShut {NoStop}%
\bibitem [{\citenamefont {Ovcharenko}\ and\ \citenamefont {Podolsk{\'y}}(2025)}]{Ovcharenko:2025cpm}%
  \BibitemOpen
  \bibfield  {author} {\bibinfo {author} {\bibfnamefont {H.}~\bibnamefont {Ovcharenko}}\ and\ \bibinfo {author} {\bibfnamefont {J.}~\bibnamefont {Podolsk{\'y}}},\ }\bibfield  {title} {\bibinfo {title} {{New class of rotating charged black holes with nonaligned electromagnetic field}},\ }\href {https://doi.org/10.1103/8wkz-th6v} {\bibfield  {journal} {\bibinfo  {journal} {Phys. Rev. D}\ }\textbf {\bibinfo {volume} {112}},\ \bibinfo {pages} {064076} (\bibinfo {year} {2025})},\ \Eprint {https://arxiv.org/abs/2508.04850} {arXiv:2508.04850 [gr-qc]} \BibitemShut {NoStop}%
\bibitem [{\citenamefont {Gray}\ \emph {et~al.}(2026)\citenamefont {Gray}, \citenamefont {Kubiznak}, \citenamefont {Ovcharenko},\ and\ \citenamefont {Podolsky}}]{Gray:2025lwy}%
  \BibitemOpen
  \bibfield  {author} {\bibinfo {author} {\bibfnamefont {F.}~\bibnamefont {Gray}}, \bibinfo {author} {\bibfnamefont {D.}~\bibnamefont {Kubiznak}}, \bibinfo {author} {\bibfnamefont {H.}~\bibnamefont {Ovcharenko}},\ and\ \bibinfo {author} {\bibfnamefont {J.}~\bibnamefont {Podolsky}},\ }\bibfield  {title} {\bibinfo {title} {{Hidden symmetries and separability structures of Ovcharenko-Podolsk{\'y} and conformal-to-Carter spacetimes}},\ }\href {https://doi.org/10.1103/8832-htpg} {\bibfield  {journal} {\bibinfo  {journal} {Phys. Rev. D}\ }\textbf {\bibinfo {volume} {113}},\ \bibinfo {pages} {044050} (\bibinfo {year} {2026})},\ \Eprint {https://arxiv.org/abs/2511.21538} {arXiv:2511.21538 [gr-qc]} \BibitemShut {NoStop}%
\bibitem [{\citenamefont {Mart\'in-Garc\'ia}()}]{Garcia:2011}%
  \BibitemOpen
  \bibfield  {author} {\bibinfo {author} {\bibfnamefont {J.}~\bibnamefont {Mart\'in-Garc\'ia}},\ }\href {http://www.xact.es/} {\bibinfo {title} {xact: Efficient tensor computer algebra}}\BibitemShut {NoStop}%
\bibitem [{\citenamefont {Collinson}(1974)}]{collinson1974existence}%
  \BibitemOpen
  \bibfield  {author} {\bibinfo {author} {\bibfnamefont {C.}~\bibnamefont {Collinson}},\ }\bibfield  {title} {\bibinfo {title} {The existence of killing tensors in empty space-times.},\ }\href@noop {} {\bibfield  {journal} {\bibinfo  {journal} {Tensor (NS) vol. 28 (1974}\ }\textbf {\bibinfo {volume} {28}},\ \bibinfo {pages} {173} (\bibinfo {year} {1974})}\BibitemShut {NoStop}%
\bibitem [{\citenamefont {Mason}\ and\ \citenamefont {Taghavi-Chabert}(2010)}]{Mason:2010zzc}%
  \BibitemOpen
  \bibfield  {author} {\bibinfo {author} {\bibfnamefont {L.~J.}\ \bibnamefont {Mason}}\ and\ \bibinfo {author} {\bibfnamefont {A.}~\bibnamefont {Taghavi-Chabert}},\ }\bibfield  {title} {\bibinfo {title} {{Killing-Yano tensors and multi-Hermitian structures}},\ }\href {https://doi.org/10.1016/j.geomphys.2010.02.008} {\bibfield  {journal} {\bibinfo  {journal} {J. Geom. Phys.}\ }\textbf {\bibinfo {volume} {60}},\ \bibinfo {pages} {907} (\bibinfo {year} {2010})},\ \Eprint {https://arxiv.org/abs/0805.3756} {arXiv:0805.3756 [math.DG]} \BibitemShut {NoStop}%
\bibitem [{\citenamefont {Kerr}(1963)}]{Kerr:1963ud}%
  \BibitemOpen
  \bibfield  {author} {\bibinfo {author} {\bibfnamefont {R.~P.}\ \bibnamefont {Kerr}},\ }\bibfield  {title} {\bibinfo {title} {{Gravitational field of a spinning mass as an example of algebraically special metrics}},\ }\href {https://doi.org/10.1103/PhysRevLett.11.237} {\bibfield  {journal} {\bibinfo  {journal} {Phys. Rev. Lett.}\ }\textbf {\bibinfo {volume} {11}},\ \bibinfo {pages} {237} (\bibinfo {year} {1963})}\BibitemShut {NoStop}%
\bibitem [{\citenamefont {Xu}\ \emph {et~al.}(2026)\citenamefont {Xu}, \citenamefont {Mai},\ and\ \citenamefont {Liang}}]{Xu:2026zgd}%
  \BibitemOpen
  \bibfield  {author} {\bibinfo {author} {\bibfnamefont {R.}~\bibnamefont {Xu}}, \bibinfo {author} {\bibfnamefont {Z.-F.}\ \bibnamefont {Mai}},\ and\ \bibinfo {author} {\bibfnamefont {D.}~\bibnamefont {Liang}},\ }\bibfield  {title} {\bibinfo {title} {{The stealth Kerr solution in the bumblebee gravity}},\ }\href {https://doi.org/10.1016/j.physletb.2026.140364} {\bibfield  {journal} {\bibinfo  {journal} {Phys. Lett. B}\ }\textbf {\bibinfo {volume} {875}},\ \bibinfo {pages} {140364} (\bibinfo {year} {2026})},\ \Eprint {https://arxiv.org/abs/2601.18809} {arXiv:2601.18809 [gr-qc]} \BibitemShut {NoStop}%
\bibitem [{\citenamefont {Newman}\ and\ \citenamefont {Janis}(1965)}]{Newman:1965tw}%
  \BibitemOpen
  \bibfield  {author} {\bibinfo {author} {\bibfnamefont {E.~T.}\ \bibnamefont {Newman}}\ and\ \bibinfo {author} {\bibfnamefont {A.~I.}\ \bibnamefont {Janis}},\ }\bibfield  {title} {\bibinfo {title} {{Note on the Kerr spinning particle metric}},\ }\href {https://doi.org/10.1063/1.1704350} {\bibfield  {journal} {\bibinfo  {journal} {J. Math. Phys.}\ }\textbf {\bibinfo {volume} {6}},\ \bibinfo {pages} {915} (\bibinfo {year} {1965})}\BibitemShut {NoStop}%
\bibitem [{\citenamefont {Taub}(1951)}]{taub1951empty}%
  \BibitemOpen
  \bibfield  {author} {\bibinfo {author} {\bibfnamefont {A.~H.}\ \bibnamefont {Taub}},\ }\bibfield  {title} {\bibinfo {title} {Empty space-times admitting a three parameter group of motions},\ }\href@noop {} {\bibfield  {journal} {\bibinfo  {journal} {Annals of Mathematics}\ }\textbf {\bibinfo {volume} {53}},\ \bibinfo {pages} {472} (\bibinfo {year} {1951})}\BibitemShut {NoStop}%
\bibitem [{\citenamefont {Newman}\ \emph {et~al.}(1963)\citenamefont {Newman}, \citenamefont {Tamburino},\ and\ \citenamefont {Unti}}]{newman1963empty}%
  \BibitemOpen
  \bibfield  {author} {\bibinfo {author} {\bibfnamefont {E.}~\bibnamefont {Newman}}, \bibinfo {author} {\bibfnamefont {L.}~\bibnamefont {Tamburino}},\ and\ \bibinfo {author} {\bibfnamefont {T.}~\bibnamefont {Unti}},\ }\bibfield  {title} {\bibinfo {title} {Empty-space generalization of the schwarzschild metric},\ }\href@noop {} {\bibfield  {journal} {\bibinfo  {journal} {Journal of Mathematical Physics}\ }\textbf {\bibinfo {volume} {4}},\ \bibinfo {pages} {915} (\bibinfo {year} {1963})}\BibitemShut {NoStop}%
\bibitem [{\citenamefont {Hawking}\ \emph {et~al.}(1999)\citenamefont {Hawking}, \citenamefont {Hunter},\ and\ \citenamefont {Page}}]{Hawking:1998ct}%
  \BibitemOpen
  \bibfield  {author} {\bibinfo {author} {\bibfnamefont {S.~W.}\ \bibnamefont {Hawking}}, \bibinfo {author} {\bibfnamefont {C.~J.}\ \bibnamefont {Hunter}},\ and\ \bibinfo {author} {\bibfnamefont {D.~N.}\ \bibnamefont {Page}},\ }\bibfield  {title} {\bibinfo {title} {{Nut charge, anti-de Sitter space and entropy}},\ }\href {https://doi.org/10.1103/PhysRevD.59.044033} {\bibfield  {journal} {\bibinfo  {journal} {Phys. Rev. D}\ }\textbf {\bibinfo {volume} {59}},\ \bibinfo {pages} {044033} (\bibinfo {year} {1999})},\ \Eprint {https://arxiv.org/abs/hep-th/9809035} {arXiv:hep-th/9809035} \BibitemShut {NoStop}%
\bibitem [{\citenamefont {Cl{\'e}ment}\ \emph {et~al.}(2015)\citenamefont {Cl{\'e}ment}, \citenamefont {Gal'tsov},\ and\ \citenamefont {Guenouche}}]{Clement:2015cxa}%
  \BibitemOpen
  \bibfield  {author} {\bibinfo {author} {\bibfnamefont {G.}~\bibnamefont {Cl{\'e}ment}}, \bibinfo {author} {\bibfnamefont {D.}~\bibnamefont {Gal'tsov}},\ and\ \bibinfo {author} {\bibfnamefont {M.}~\bibnamefont {Guenouche}},\ }\bibfield  {title} {\bibinfo {title} {{Rehabilitating space-times with NUTs}},\ }\href {https://doi.org/10.1016/j.physletb.2015.09.074} {\bibfield  {journal} {\bibinfo  {journal} {Phys. Lett. B}\ }\textbf {\bibinfo {volume} {750}},\ \bibinfo {pages} {591} (\bibinfo {year} {2015})},\ \Eprint {https://arxiv.org/abs/1508.07622} {arXiv:1508.07622 [hep-th]} \BibitemShut {NoStop}%
\bibitem [{\citenamefont {Gibbons}\ and\ \citenamefont {Ruback}(1988)}]{Gibbons:1987sp}%
  \BibitemOpen
  \bibfield  {author} {\bibinfo {author} {\bibfnamefont {G.~W.}\ \bibnamefont {Gibbons}}\ and\ \bibinfo {author} {\bibfnamefont {P.~J.}\ \bibnamefont {Ruback}},\ }\bibfield  {title} {\bibinfo {title} {{The Hidden Symmetries of Multicenter Metrics}},\ }\href {https://doi.org/10.1007/BF01466773} {\bibfield  {journal} {\bibinfo  {journal} {Commun. Math. Phys.}\ }\textbf {\bibinfo {volume} {115}},\ \bibinfo {pages} {267} (\bibinfo {year} {1988})}\BibitemShut {NoStop}%
\bibitem [{\citenamefont {Jezierski}\ and\ \citenamefont {Lukasik}(2007)}]{Jezierski:2006fw}%
  \BibitemOpen
  \bibfield  {author} {\bibinfo {author} {\bibfnamefont {J.}~\bibnamefont {Jezierski}}\ and\ \bibinfo {author} {\bibfnamefont {M.}~\bibnamefont {Lukasik}},\ }\bibfield  {title} {\bibinfo {title} {{Conformal Yano-Killing tensors for the Taub-NUT metric}},\ }\href {https://doi.org/10.1088/0264-9381/24/5/015} {\bibfield  {journal} {\bibinfo  {journal} {Class. Quant. Grav.}\ }\textbf {\bibinfo {volume} {24}},\ \bibinfo {pages} {1331} (\bibinfo {year} {2007})},\ \Eprint {https://arxiv.org/abs/gr-qc/0610090} {arXiv:gr-qc/0610090} \BibitemShut {NoStop}%
\bibitem [{\citenamefont {Arias}\ \emph {et~al.}()\citenamefont {Arias}, \citenamefont {Kop{\v c}an}, \citenamefont {Kubiznak},\ and\ \citenamefont {Mili{\v c}ka}}]{Arias:2016}%
  \BibitemOpen
  \bibfield  {author} {\bibinfo {author} {\bibfnamefont {C.}~\bibnamefont {Arias}}, \bibinfo {author} {\bibfnamefont {D.}~\bibnamefont {Kop{\v c}an}}, \bibinfo {author} {\bibfnamefont {D.}~\bibnamefont {Kubiznak}},\ and\ \bibinfo {author} {\bibfnamefont {M.}~\bibnamefont {Mili{\v c}ka}},\ }\bibfield  {title} {\bibinfo {title} {{On Generalized Killing Tensors, to be published (2026)}},\ }\href@noop {} {\ }\BibitemShut {NoStop}%
\bibitem [{\citenamefont {Dietz}\ and\ \citenamefont {R{\"u}diger}(1981)}]{dietz1981space}%
  \BibitemOpen
  \bibfield  {author} {\bibinfo {author} {\bibfnamefont {W.}~\bibnamefont {Dietz}}\ and\ \bibinfo {author} {\bibfnamefont {R.}~\bibnamefont {R{\"u}diger}},\ }\bibfield  {title} {\bibinfo {title} {Space--times admitting killing--yano tensors. i},\ }\href@noop {} {\bibfield  {journal} {\bibinfo  {journal} {Proceedings of the Royal Society of London. A. Mathematical and Physical Sciences}\ }\textbf {\bibinfo {volume} {375}},\ \bibinfo {pages} {361} (\bibinfo {year} {1981})}\BibitemShut {NoStop}%
\bibitem [{\citenamefont {Myers}\ and\ \citenamefont {Perry}(1986)}]{Myers:1986un}%
  \BibitemOpen
  \bibfield  {author} {\bibinfo {author} {\bibfnamefont {R.~C.}\ \bibnamefont {Myers}}\ and\ \bibinfo {author} {\bibfnamefont {M.~J.}\ \bibnamefont {Perry}},\ }\bibfield  {title} {\bibinfo {title} {{Black Holes in Higher Dimensional Space-Times}},\ }\href {https://doi.org/10.1016/0003-4916(86)90186-7} {\bibfield  {journal} {\bibinfo  {journal} {Annals Phys.}\ }\textbf {\bibinfo {volume} {172}},\ \bibinfo {pages} {304} (\bibinfo {year} {1986})}\BibitemShut {NoStop}%
\bibitem [{\citenamefont {Kubiznak}\ \emph {et~al.}(2009)\citenamefont {Kubiznak}, \citenamefont {Kunduri},\ and\ \citenamefont {Yasui}}]{Kubiznak:2009qi}%
  \BibitemOpen
  \bibfield  {author} {\bibinfo {author} {\bibfnamefont {D.}~\bibnamefont {Kubiznak}}, \bibinfo {author} {\bibfnamefont {H.~K.}\ \bibnamefont {Kunduri}},\ and\ \bibinfo {author} {\bibfnamefont {Y.}~\bibnamefont {Yasui}},\ }\bibfield  {title} {\bibinfo {title} {{Generalized Killing-Yano equations in D=5 gauged supergravity}},\ }\href {https://doi.org/10.1016/j.physletb.2009.06.037} {\bibfield  {journal} {\bibinfo  {journal} {Phys. Lett. B}\ }\textbf {\bibinfo {volume} {678}},\ \bibinfo {pages} {240} (\bibinfo {year} {2009})},\ \Eprint {https://arxiv.org/abs/0905.0722} {arXiv:0905.0722 [hep-th]} \BibitemShut {NoStop}%
\bibitem [{\citenamefont {Houri}\ \emph {et~al.}(2010{\natexlab{a}})\citenamefont {Houri}, \citenamefont {Kubiznak}, \citenamefont {Warnick},\ and\ \citenamefont {Yasui}}]{Houri:2010qc}%
  \BibitemOpen
  \bibfield  {author} {\bibinfo {author} {\bibfnamefont {T.}~\bibnamefont {Houri}}, \bibinfo {author} {\bibfnamefont {D.}~\bibnamefont {Kubiznak}}, \bibinfo {author} {\bibfnamefont {C.}~\bibnamefont {Warnick}},\ and\ \bibinfo {author} {\bibfnamefont {Y.}~\bibnamefont {Yasui}},\ }\bibfield  {title} {\bibinfo {title} {{Symmetries of the Dirac Operator with Skew-Symmetric Torsion}},\ }\href {https://doi.org/10.1088/0264-9381/27/18/185019} {\bibfield  {journal} {\bibinfo  {journal} {Class. Quant. Grav.}\ }\textbf {\bibinfo {volume} {27}},\ \bibinfo {pages} {185019} (\bibinfo {year} {2010}{\natexlab{a}})},\ \Eprint {https://arxiv.org/abs/1002.3616} {arXiv:1002.3616 [hep-th]} \BibitemShut {NoStop}%
\bibitem [{\citenamefont {Houri}\ \emph {et~al.}(2010{\natexlab{b}})\citenamefont {Houri}, \citenamefont {Kubiznak}, \citenamefont {Warnick},\ and\ \citenamefont {Yasui}}]{Houri:2010fr}%
  \BibitemOpen
  \bibfield  {author} {\bibinfo {author} {\bibfnamefont {T.}~\bibnamefont {Houri}}, \bibinfo {author} {\bibfnamefont {D.}~\bibnamefont {Kubiznak}}, \bibinfo {author} {\bibfnamefont {C.~M.}\ \bibnamefont {Warnick}},\ and\ \bibinfo {author} {\bibfnamefont {Y.}~\bibnamefont {Yasui}},\ }\bibfield  {title} {\bibinfo {title} {{Generalized hidden symmetries and the Kerr-Sen black hole}},\ }\href {https://doi.org/10.1007/JHEP07(2010)055} {\bibfield  {journal} {\bibinfo  {journal} {JHEP}\ }\textbf {\bibinfo {volume} {07}},\ \bibinfo {pages} {055}},\ \Eprint {https://arxiv.org/abs/1004.1032} {arXiv:1004.1032 [hep-th]} \BibitemShut {NoStop}%
\bibitem [{\citenamefont {Collinson}\ and\ \citenamefont {Howarth}(2000)}]{collinson2000generalized}%
  \BibitemOpen
  \bibfield  {author} {\bibinfo {author} {\bibfnamefont {C.}~\bibnamefont {Collinson}}\ and\ \bibinfo {author} {\bibfnamefont {L.}~\bibnamefont {Howarth}},\ }\bibfield  {title} {\bibinfo {title} {Generalized killing tensors},\ }\href@noop {} {\bibfield  {journal} {\bibinfo  {journal} {General Relativity and Gravitation}\ }\textbf {\bibinfo {volume} {32}},\ \bibinfo {pages} {1767} (\bibinfo {year} {2000})}\BibitemShut {NoStop}%
\bibitem [{\citenamefont {Houri}\ \emph {et~al.}(2018)\citenamefont {Houri}, \citenamefont {Tomoda},\ and\ \citenamefont {Yasui}}]{Houri:2017tlk}%
  \BibitemOpen
  \bibfield  {author} {\bibinfo {author} {\bibfnamefont {T.}~\bibnamefont {Houri}}, \bibinfo {author} {\bibfnamefont {K.}~\bibnamefont {Tomoda}},\ and\ \bibinfo {author} {\bibfnamefont {Y.}~\bibnamefont {Yasui}},\ }\bibfield  {title} {\bibinfo {title} {{On integrability of the Killing equation}},\ }\href {https://doi.org/10.1088/1361-6382/aaa4e7} {\bibfield  {journal} {\bibinfo  {journal} {Class. Quant. Grav.}\ }\textbf {\bibinfo {volume} {35}},\ \bibinfo {pages} {075014} (\bibinfo {year} {2018})},\ \Eprint {https://arxiv.org/abs/1704.02074} {arXiv:1704.02074 [gr-qc]} \BibitemShut {NoStop}%
\bibitem [{\citenamefont {Carter}(1968{\natexlab{a}})}]{carter1968new}%
  \BibitemOpen
  \bibfield  {author} {\bibinfo {author} {\bibfnamefont {B.}~\bibnamefont {Carter}},\ }\bibfield  {title} {\bibinfo {title} {A new family of einstein spaces},\ }\href@noop {} {\bibfield  {journal} {\bibinfo  {journal} {Physics Letters A}\ }\textbf {\bibinfo {volume} {26}},\ \bibinfo {pages} {399} (\bibinfo {year} {1968}{\natexlab{a}})}\BibitemShut {NoStop}%
\bibitem [{\citenamefont {Carter}(1968{\natexlab{b}})}]{carter1968hamilton}%
  \BibitemOpen
  \bibfield  {author} {\bibinfo {author} {\bibfnamefont {B.}~\bibnamefont {Carter}},\ }\bibfield  {title} {\bibinfo {title} {Hamilton-jacobi and schrodinger separable solutions of einstein’s equations},\ }\href@noop {} {\bibfield  {journal} {\bibinfo  {journal} {Communications in Mathematical Physics}\ }\textbf {\bibinfo {volume} {10}},\ \bibinfo {pages} {280} (\bibinfo {year} {1968}{\natexlab{b}})}\BibitemShut {NoStop}%
\bibitem [{\citenamefont {Griffiths}\ and\ \citenamefont {Podolsky}(2006)}]{Griffiths:2005qp}%
  \BibitemOpen
  \bibfield  {author} {\bibinfo {author} {\bibfnamefont {J.~B.}\ \bibnamefont {Griffiths}}\ and\ \bibinfo {author} {\bibfnamefont {J.}~\bibnamefont {Podolsky}},\ }\bibfield  {title} {\bibinfo {title} {{A New look at the Plebanski-Demianski family of solutions}},\ }\href {https://doi.org/10.1142/S0218271806007742} {\bibfield  {journal} {\bibinfo  {journal} {Int. J. Mod. Phys. D}\ }\textbf {\bibinfo {volume} {15}},\ \bibinfo {pages} {335} (\bibinfo {year} {2006})},\ \Eprint {https://arxiv.org/abs/gr-qc/0511091} {arXiv:gr-qc/0511091} \BibitemShut {NoStop}%
\bibitem [{\citenamefont {Ovcharenko}\ \emph {et~al.}(2025{\natexlab{a}})\citenamefont {Ovcharenko}, \citenamefont {Podolsky},\ and\ \citenamefont {Astorino}}]{Ovcharenko:2024yyu}%
  \BibitemOpen
  \bibfield  {author} {\bibinfo {author} {\bibfnamefont {H.}~\bibnamefont {Ovcharenko}}, \bibinfo {author} {\bibfnamefont {J.}~\bibnamefont {Podolsky}},\ and\ \bibinfo {author} {\bibfnamefont {M.}~\bibnamefont {Astorino}},\ }\bibfield  {title} {\bibinfo {title} {{Black holes of type D revisited: Relating their various metric forms}},\ }\href {https://doi.org/10.1103/PhysRevD.111.024038} {\bibfield  {journal} {\bibinfo  {journal} {Phys. Rev. D}\ }\textbf {\bibinfo {volume} {111}},\ \bibinfo {pages} {024038} (\bibinfo {year} {2025}{\natexlab{a}})},\ \Eprint {https://arxiv.org/abs/2409.02308} {arXiv:2409.02308 [gr-qc]} \BibitemShut {NoStop}%
\bibitem [{\citenamefont {Ovcharenko}\ \emph {et~al.}(2025{\natexlab{b}})\citenamefont {Ovcharenko}, \citenamefont {Podolsky},\ and\ \citenamefont {Astorino}}]{Ovcharenko:2025fxg}%
  \BibitemOpen
  \bibfield  {author} {\bibinfo {author} {\bibfnamefont {H.}~\bibnamefont {Ovcharenko}}, \bibinfo {author} {\bibfnamefont {J.}~\bibnamefont {Podolsky}},\ and\ \bibinfo {author} {\bibfnamefont {M.}~\bibnamefont {Astorino}},\ }\bibfield  {title} {\bibinfo {title} {{Revisiting black holes of algebraic type D with a cosmological constant}},\ }\href {https://doi.org/10.1103/PhysRevD.111.084016} {\bibfield  {journal} {\bibinfo  {journal} {Phys. Rev. D}\ }\textbf {\bibinfo {volume} {111}},\ \bibinfo {pages} {084016} (\bibinfo {year} {2025}{\natexlab{b}})},\ \Eprint {https://arxiv.org/abs/2501.07537} {arXiv:2501.07537 [gr-qc]} \BibitemShut {NoStop}%
\bibitem [{\citenamefont {Kubiznak}\ and\ \citenamefont {Krtous}(2007)}]{Kubiznak:2007kh}%
  \BibitemOpen
  \bibfield  {author} {\bibinfo {author} {\bibfnamefont {D.}~\bibnamefont {Kubiznak}}\ and\ \bibinfo {author} {\bibfnamefont {P.}~\bibnamefont {Krtous}},\ }\bibfield  {title} {\bibinfo {title} {{On conformal Killing-Yano tensors for Plebanski-Demianski family of solutions}},\ }\href {https://doi.org/10.1103/PhysRevD.76.084036} {\bibfield  {journal} {\bibinfo  {journal} {Phys. Rev. D}\ }\textbf {\bibinfo {volume} {76}},\ \bibinfo {pages} {084036} (\bibinfo {year} {2007})},\ \Eprint {https://arxiv.org/abs/0707.0409} {arXiv:0707.0409 [gr-qc]} \BibitemShut {NoStop}%
\end{thebibliography}

%

\end{document}